\documentclass[10pt,aps,prd,twocolumn,showpacs,nofootinbib,amsmath,amssymb,
  superscriptaddress,floatfix]{revtex4-2}

\usepackage{graphicx}
\graphicspath{{./}}
\usepackage[dvipsnames]{xcolor}
\newcommand{\rev}[1]{{\color{black}#1}}
\usepackage{bm}
\usepackage{booktabs}
\usepackage{multirow}
\usepackage[expansion=false]{microtype}
\usepackage{placeins}
\usepackage{hyperref}
\hypersetup{colorlinks=true,linkcolor=NavyBlue,citecolor=NavyBlue,
  urlcolor=NavyBlue,breaklinks=true,
  pdftitle={Two-Fluid Radial Stability of Dark-Matter-Admixed Compact Objects: Dark Compact Planets, Dark White Dwarfs, Dark-Matter-Admixed Neutron Stars, and Dark Oysters},
  pdfauthor={Jose C. Jimenez, Oscar Pastor, Laura Tolos},
  pdfkeywords={two-fluid stars, radial stability, fermionic dark matter}}
  \usepackage{soul}

\newcommand{\OM}{\mathrm{OM}}
\newcommand{\DM}{\mathrm{DM}}
\newcommand{\DCP}{\mathrm{DCP}}
\newcommand{\DWD}{\mathrm{DWD}}
\newcommand{\DMANS}{\mathrm{\color{black}DANS}}
\newcommand{\DOY}{\mathrm{DOY}}
\newcommand{\Msun}{M_{\odot}}
\newcommand{\Mjup}{M_{\mathrm{Jupiter}}}
\newcommand{\Mearth}{M_{\oplus}}
\newcommand{\dd}{\mathrm{d}}
\newcommand{\articlegraphic}[2][]{\includegraphics[#1]{#2}}

\begin{document}
\raggedbottom

\title{Two-Fluid Radial Stability of Dark-Matter-Admixed Compact Objects:\\
Dark Compact Planets, Dark White Dwarfs, Dark Neutron Stars, and Dark Oysters}

\author{Jos\'e C. Jim\'enez}
\affiliation{Department of Astrophysics, The Brazilian Center for Research
in Physics (CBPF), Rua Dr. Xavier Sigaud, 150, URCA,
Rio de Janeiro CEP 22210-180, RJ, Brazil}
\affiliation{Center for Nuclear Research, Department of Physics, Kent State University, Kent, OH 44242 USA}

\author{Oscar Pastor}
\affiliation{Grupo de Astrof\'isica Te\'orica y Part\'iculas, Escuela Profesional de Física, Universidad Nacional de San Agust\'in de Arequipa,
Av. Independencia S/N, Arequipa 04000, Peru}

\author{Laura Tolos}
\affiliation{Institute of Space Sciences (ICE, CSIC), Campus UAB,
Carrer de Can Magrans, 08193 Barcelona, Spain}
\affiliation{Institut d'Estudis Espacials de Catalunya (IEEC),
08860 Castelldefels, Barcelona, Spain}

\date{\today}

\begin{abstract}
We investigate the equilibrium structure and stability against radial oscillations of a variety of two-fluid compact objects containing ordinary baryonic matter admixed with cold dark matter coupled through the background spacetime. For the ordinary fluid, we employ a QCD equation of state compatible with chiral effective field theory at nuclear densities (supplemented by the BPS and Negele-Vautherin models at extremely low densities) and with perturbative QCD in the high-density regime while also satisfying also current astrophysical data. For the dark fluid, we assume self-interacting fermionic matter with particle masses of $m_f=$ 2, 5, 10, and 100 GeV and two self-interaction strengths of $y=0.1$ and $10^3$. By scanning a large range of central pressures for both fluids, we identify the presence of dark compact planets, dark white dwarfs, dark neutron stars and the so-called dark oysters. Importantly, these objects are not only in two-fluid hydrostatic equilibrium, but they are also stable against two-fluid radial oscillations. We find that the relative representation of each class in the radially stable sample is strongly sensitive to the combinations of $m_f$ and $y$, with some combinations entirely suppressing some of these object classes within the sampled domain while favoring others. In addition to these classes, strong self-repulsion supports stable, dark-matter-dominated branches with compact ordinary-matter cores and extended dark halos, approaching the self-interacting fermion-star limit. \rev{Finally, we calculate the gravitational redshift at the ordinary-matter surface, $z_{\rm OM}$, including the contribution of extended dark-matter halos for which we found radially stable configurations with $z_{\OM}>2$.}
\end{abstract}

\maketitle
\section{Introduction}
\label{sec:introduction}

Although cosmological and galactic observations provide compelling evidence in favor of the dark matter (DM) hypothesis, its microscopic composition is still unknown~\cite{Planck2020}. In order to get insights about these potential dark particles and their interaction channels within models beyond the standard model of particle physics, several experiments and observatories on Earth and in space aim to start constraining their masses and dynamics when interacting with ordinary matter (e.g., atoms, ions, crystals) taking advantage of new quantum sensors and other innovative detectors \cite{Bertone:2019irm,Akerib:2022ort,Essig:2022dfa}.

Interestingly, neutron stars (NSs), being remnants of massive stars undergoing core collapse \cite{Glendenning:2000,Schaffner:2020} and unique astrophysical laboratories to probe the microphysics of cold and dense quantum chromodynamics (QCD) \cite{Sorensen:2023zkk}, might accumulate DM over the hundreds of millions of years of their lives, mainly those located close to the galactic centers, where the DM density is expected to be higher and sizeable aggregation might nontrivially affect their stellar observable properties like masses, radii and tidal deformabilities (see Refs. \cite{Bertone:2007ae,Massey:2010hh,Kouvaris:2010jy,Ellis:2018bkr,Collier:2022cpr} and references therein for further details).

Remarkably, in recent years, many advances have been made by using NSs as an arena to test DM features employing current knowledge from their structural and dynamical properties. For instance, some works consider DM effects on gravitational-wave \cite{Nelson:2018xtr,Karkevandi:2021ygv} and electromagnetic \cite{Miao:2022rqj,Shakeri:2022dwg} signals when passing through dark halos surrounding NSs, while other works study the effects of interactions between DM and ordinary baryonic matter (OM) in the total equation of state (EoS) \cite{Das:2020vng}. DM might also provide smoking-gun signatures to identify the presence of exotic objects not fully identifiable just by gravitational waves, like in the GW190814  merger event \cite{Das:2021yny}. Naturally, DM could affect the properties of rapidly rotating NSs, such as their moments of inertia, equatorial radius and universal relations even if added in small amounts \cite{Konstantinou:2024ynd}. DM-induced changes in hybrid-star structure, including quark cores at low total masses and extended dark halos, were studied in Ref. \cite{Biesdorf2025}. In contrast, hybrid NSs with smooth transitions were also explored showing nontrivial modifications of their mass-radius relations comparable with upcoming data \cite{Wei:2026gcc} (see Refs. \cite{Grippa:2024ach,Jockel:2023rrm} for recent reviews). Some works also considered the possibility of DM in protoneutron star formation with rich hadronic composition in the stationary \cite{Issifu:2024htq,Issifu:2025jac} and rotating \cite{Issifu:2025gsq} cases.

However, it is equally likely that DM accumulates around other well-known OM objects, i.e. white dwarfs and planets, which have much lower central pressures compared to those of NSs \cite{Schaffner:2020}. On one hand, these objects are complex systems with rich compositions affected by their progenitors, formation processes, different lifetimes and environmental conditions, e.g. white dwarfs might be composed of carbon or oxygen ions instead of just an electron sea while planets might be mostly rocky or gaseous \cite{Schaffner:2020}. On the other hand, these objects can be theoretically predicted having similar masses and radii directly from the low-density sector of the cold and dense nuclear matter EoS after solving the corresponding structure equations. As expected, there exist some quantitative differences related to the aforementioned evolutionary details of each object affecting their masses and radii, but the basic microphysics is the same as those obtained from the nuclear matter EoS. Of course, the ideal scenario would require a unified EoS for matter spanning all possible densities from hydrogen (for gaseous planets) to the cold quark matter phase. However, this is not necessarily a good approach to study these objects since the density scales are quite separated by several orders of magnitude. In that sense, from now on, we refer mostly to `planets' and `white dwarfs' as those merely obtained from the nuclear EoS having their characteristic sizes and masses similar to those observed in nature and leaving aside their evolutionary chemical and plasma complexity, as already done in Refs. \cite{Tolos2015,Dengler2022}. Either way, in Sec. \ref{sec:results}, we discuss modifications caused by using a realistic EoS for cold white dwarfs.

In this work, we explore the effects that fermionic\footnote{Recently, the LIGO-Virgo-KAGRA detectors \cite{LIGOScientific:2025ttj} reported no evidence of some types of bosonic DM \cite{Karkevandi:2024vov} after analyzing data of their 4th observing run and put stringent upper limits on the OM-DM couplings. Thus, fermionic DM still an attractive prospect.} DM has on low-pressure (quasi)stellar objects which we dub as dark-matter admixed compact objects (DMACOs). In order to do so, we use the EoS-I of Annala et al. \cite{Annala2020} valid at all baryon densities\footnote{See Appendix C of the first arXiv version of this article.} satisfying also modern astrophysical constraints while NSs close to the maximum mass have quark cores through a crossover transition. Notice that this EoS-I contains the BPS and Negele-Vautherin EoS at extremely low pressures usually employed for the crusts but now relevant for the DMACOs As it is known, using this type of EoS for a wide range of densities manifests as in Figs. 7.9--7.11 from Ref. \cite{Schaffner:2020} and Fig. 12.10 from Ref. \cite{Glendenning:2000} after solving the Tolman-Oppenheimer-Volkoff  (TOV) equations for stellar structure\footnote{Although it is true that one can use the Newtonian limit of these equations for white dwarfs and planets, we prefer to employ the general-relativistic version in order to leave the code ready to be extended to the two-fluid case, where the dark fluid introduces general-relativistic changes to the whole two-fluid admixed star.}, where one can distinguish the {\it planet}, {\it white-dwarf} and NS branches. 

Unfortunately, the usual static criterion of stability $dM/dp_c\geq 0$ (for a given central pressure $p_c$) \cite{Harrison:1965abs,Glendenning:2000} is necessary, but not sufficient to distinguish dynamically stable from unstable stars. In this case, at low central pressures (below the white-dwarf sector) and when approaching the NS branch, some stars are unstable ($dM/dp_c < 0$) while others satisfy $dM/dp_c > 0$. This was problematic since these apparently stable objects were undiscovered by astronomers. It was just after the careful analysis of Refs. \cite{Bardeen1966,Meltzer:1966abs} in 1966 based upon the formalism of adiabatic linear radial oscillations developed by S. Chandrasekhar \cite{Chandrasekhar1964} in 1964 that these studies established which planetary, white-dwarf, and NS branches are radially stable in general relativity, something which was in full agreement with natural objects discovered by that time having masses and radii similar to those predicted by the nuclear EoS. Although similar discussions can be shown based on the behavior on the extrema of their radii $R$, $dR/dp_c$, only the Chandrasekhar perspective determines reliably which configurations are fully stable. 

From this perspective, if we view planets, white dwarfs present in nature as those objects which are necessarily radially stable obtainable from the TOV solutions from the dense QCD matter EoS, we are immediately led to ask if adding DM to the stellar objects obtained from the low-pressure EoS (including the radially unstable regions) could still allow the appearance of radially stable configurations with the potential to be discovered in nature in the upcoming future with multimessenger data. 

By adopting that path of reasoning, we can employ the two-fluid approach: OM interacts with DM just through the gravitational interaction, being each one of them considered as perfect fluids with their own EoS and number-conservation equations~\cite{Goldman2013,Leung2011}. Within this formalism, a hydrostatically equilibrated configuration is specified by the two central pressures, $p_{\OM}^c$ and $p_{\DM}^c$. Depending on the relative content and spatial extent of the components, DM can form a compact core, an extended halo, or dominate the overall configuration.

For the particular case of white dwarfs and NSs admixed with fermionic DM \cite{Narain2006}, past studies ~\cite{Goldman2013,Leung2011,Leung2013,Deliyergiyev2019,Barbat2024} had already explored this possibility, although focusing only on the two-fluid stellar structure. It was not until the work of Ref. \cite{Tolos2015} that
by considering fermionic DM admixed with NS at all densities produce {\it dark compact planets} (DCPs) with much smaller radii than OM planets, but with
comparable mass and {\it dark white dwarfs} (DWDs) having also lower masses but having similar sizes~\cite{Tolos2015}. In a subsequent work, the same authors \cite{Dengler2022} studied their tidal response for weak and strong
self-interactions over a broad range of dark-particle masses.

As already explained, hydrostatic equilibrium alone does not establish dynamical stability. In the one-fluid problem, the onset of radial instability is identified by a vanishing fundamental squared frequency~\cite{Chandrasekhar1964,Bardeen1966}. With two
independently conserved fluids, the displacements satisfy a coupled
pulsation problem and stability is determined by its lowest global
eigenvalue, not by either component in isolation~\cite{Kain2020}.
While Ref. \cite{Leung2012} calculated radial modes of fermionic-DM-admixed NSs using the less-known general-relativistic Carter's two-fluid approach in a very simplified and unclear manner, Kain~\cite{Kain2020,Kain2021} developed the relativistic multifluid formalism generalizing the one-fluid Chandrasekhar equations and mapped fundamental-mode frequencies over extended central-pressure domains in a complete manner. Related work examined DM-admixed quark stars~\cite{JimenezFraga2022}, nonlinear dynamical evolution of these two-fluid oscillations~\cite{Gleason2022}, and the mathematical formalization for this double eigenvalue problem~\cite{Caballero2024}.

Inspired by the work of Ref. \cite{Henriques1990}, the author of Ref. \cite{Kain2021} proposed an equilibrium-based alternative to construct a generalized turning-point criterion using the two conserved fluid numbers. He found a perfect agreement when calculating the stability curve employing the two-fluid radial oscillation equation and the two-fluid turning-point method (based upon computing the extrema of $M$ and the particle numbers) when applied to DM admixed NSs. This was also verified using a more formal approach \cite{Caballero2024} and using realistic (tabulated) EoS~\cite{KumarSotani2025}. Besides, for regular equilibrium
families of separately conserved, isentropic barotropes coupled only by
gravity, Ref. \cite{Chen2026} proved that loss of rank of the particle-number
map is equivalent to a physical radial zero mode. Thus, identifying the crossing mode and the stable side still requires information about the sign of the squared eigenvalues. This is also because internal critical curves can correspond to excited-mode zeros inside an already unstable domain without contradicting the turning-point criterion. In that sense, the immediate and blind use of the generalized turning-point technique might induce an incorrect classification of the stability curves which might be connected to the change of signs of the overtone eigenfrequencies rather than the fundamental mode, which uniquely determines two-fluid stable configurations.

This motivates a comprehensive calculation of the two-fluid radial oscillation frequencies for these DM admixed planets and white dwarfs, which in the past have been studied through their equilibrium structure and tidal deformabilities~\cite{Leung2013,Tolos2015,Dengler2022}. Besides, the radial stability of DM admixed white dwarfs was studied in the simplified one-fluid approach in Refs.~\cite{Leung2013,Carvalho2025}. Instead, we solve the coupled eigenvalue problem retaining independent fluid displacements to determine the lowest global mode, which classifies properly the structural stability. Mapping this spectrum across the low-pressure diagram allows us to identify stable configurations beyond simply examining the mass--radius relations.

The remainder of the paper is organized as follows. Section~\ref{sec:setup}
introduces the two-fluid TOV equations, the equations of state from Annala \textit{et al.}~\cite{Annala2020} for OM and a self-interacting
degenerate Fermi gas~\cite{Narain2006,Tolos2015,Dengler2022} for DM with $m_f=2$, 5, 10, and $100~\mathrm{GeV}$ with weak and strong self-repulsion, $y=0.1$ and $10^3$. Besides, the static and dynamical stability criteria are briefly summarized as well as the compact-object classification. Section~\ref{sec:results} presents the obtained mass--radius relations, DM
fractions, and calculation of the static critical curves, the two-fluid stable low-pressure domains, the radial-stability maps for their fundamental radial frequencies  and the resulting radially stable compact-object families. 
We use $G=c=\hbar=1$ units unless otherwise stated. 

\section{The Two-fluid Approach}
\label{sec:setup}
\subsection{Hydrostatic Equilibrium}
\label{sec:tov}

As occurs with the one-fluid TOV equations
\cite{Tolman1939,OppenheimerVolkoff1939} for a static and spherically
symmetric spacetime, one has the two-fluid generalization given by \cite{Kain2020,Kain2021}
\begin{align}
\frac{\dd m_X}{\dd r} &=4\pi r^2\epsilon_X,
\label{eq:tov_mass}\\
\frac{\dd p_X}{\dd r} &=-\left(\epsilon_X+p_X\right)
\frac{m+4\pi r^3 p}{r(r-2m)},
\label{eq:tov_pressure}
\end{align}
where $X\in\{\OM,\DM\}$ denoting the OM or DM fluid, $\epsilon_X(r)$ as the energy densities, $p_X(r)$ as the corresponding pressures, $m_X(r)$ as the gravitational mass enclosed by a given fluid $X$ within a radius $r$,  $m=m_{\OM}+m_{\DM}$ the total enclosed gravitational mass, and $p=p_{\OM}+p_{\DM}$ as the total pressure. Each fluid ends at its own surface $R_X$, defined by $p_X(R_X)=0$. After the inner component vanishes, the outer component continues as a one-fluid envelope. The fluid's gravitational masses are $M_X=m_X(R_X)$, the outer radius is $R=\max(R_{\OM},R_{\DM})$, and the total gravitational mass is $M=M_{\OM}+M_{\DM}=m(R)$. For later purposes, we define the dimensionless DM mass fraction as $F_{\DM}\equiv M_{\DM}/M$. Its numerical value in percent is $100F_{\DM}$; colorbars labeled $F_{\DM}$ [\%] use this percentage convention. We distinguish a DM core if $R_{\DM}<R_{\OM}$ or a DM halo if $R_{\DM}>R_{\OM}$.

It is also necessary to compute the total fluid number (interpreted as a conserved Noether charge) from \cite{Kain2020,Kain2021}
\begin{equation}
N_X=4\pi\int_0^{\rev{R_X}}\frac{\rev{n_X}(r)r^2}{\sqrt{1-2m(r)/r}}\,\dd r, 
\label{eq:particle_number}
\end{equation}
For each fluid, we obtain a number density consistent with the interpolated barotrope by integrating \rev{$\dd\ln n_X=\dd\epsilon_X/(\epsilon_X+p_X)$}~\cite{Kain2020,Kain2021}, where the normalization constants will be unimportant. Notice that one must use the EoS for OM/DM to further construct this $n_X$ instead of just taking the particle density from the EoS, like the baryon density for OM, \rev{as done in Ref. \cite{Hippert2023}}.

\subsection{OM and DM Equations of State}
\label{sec:eos}

The ordinary fluid is described by the supplied tabulation of the representative EoS-I of Annala et al.~\cite{Annala2020}. Its core construction uses chiral effective field theory at nuclear densities while constrained by perturbative QCD at high densities, together with the astrophysical constraints adopted in that work. \rev{Notice that although all our DCPs and DWDs use the BPS and Negele-Vautherin EoS \cite{Hebeler:2013nza} for extremely low baryon densities, we} examine a different white-dwarf matter model in Sec.~\ref{sec:dwd_eos_test}.

The DM fluid is modeled as a cold, self-interacting degenerate Fermi gas \cite{Narain2006,Tolos2015}. Its two free microscopic parameters are
the fermion mass $m_f$ and the interaction strength $y=m_f/m_I$, where
$m_I$ is the interaction mass scale. The repulsive contribution stiffens the DM EoS. We consider $m_f=2$, 5, 10 and $100~\mathrm{GeV}$ with $y=0.1$ and $10^3$ as parameters which are in agreement with astrophysical and cosmological DM models \cite{KumarSotani2025}.

\subsection{Static Method: Generalized Turning Point}
\label{sec:stability}

After choosing the EoS for each fluid, a given stellar two-fluid configuration having $M$, $N_{\rm OM}$ and $N_{\rm DM}$ can be parametrized by the pair $q=(p_{\OM}^c,p_{\DM}^c)$ corresponding to the central pressures of each fluid. The generalized turning-point condition identifies critical configurations by requiring a nonzero tangent $\mathbf t$ for which \cite{Henriques1990,Kain2021}
\begin{equation}
D_{\mathbf t}M=D_{\mathbf t}N_{\OM}=D_{\mathbf t}N_{\DM}=0,
\label{eq:turning_point}
\end{equation}
where $D_{\mathbf t}$ denotes the directional derivative along the
equilibrium family. Equivalently, a particle-number-preserving tangent exists when
\begin{equation}
J\equiv
\det\left[
\frac{\partial(N_{\OM},N_{\DM})}
     {\partial(p_{\OM}^c,p_{\DM}^c)}
\right]=0.
\label{eq:jacobian}
\end{equation}
Then, if the condition of Eq. (\ref{eq:jacobian}) is satisfied, Eq. (\ref{eq:turning_point}) is immediately satisfied for the stationary-mass condition. However, one should keep in mind that for a given pair of central pressures, this rank condition generally defines critical curves rather than isolated extrema. Under the regularity and thermodynamic hypotheses of
Ref.~\cite{Chen2026}, rank loss is equivalent to a null radial mode, but
it does not identify which particular mode vanishes. Thus, it becomes mandatory to further solve the two-fluid radial oscillation equation to find the ordered spectrum so as to distinguish the fundamental-mode boundary from other curves where excited modes change their sign.

\subsection{Dynamic Method: Two-fluid Radial Frequencies}
\label{sec:dynamical_stability}

The criterion of dynamical stability of the two-fluid TOV solutions follows directly from the coupled pulsation problem of Ref.~\cite{Kain2020}. Each fluid $X$ has a Lagrangian radial displacement $\xi_X(t,r)=\xi_X(r)e^{i\omega t}$, with a common angular eigenfrequency $\omega$. In terms of $\zeta_X=r^2e^{-\nu/2}\xi_X$, the component equations have the schematic structure
\begin{multline}
\rev{(\Pi_X\zeta_X')'}
+\left(Q_X+\omega^2W_X\right)\zeta_X\\
+\sum_{Y\ne X}
\Bigl[A_{XY}\zeta_Y+B_{XY}(\zeta_Y-\zeta_X)\\
+C_{XY}(\zeta_Y'-\zeta_X')\Bigr]=0,
\label{eq:pulsation}
\end{multline}
where a prime denotes \rev{differentiation with respect to $r$}, the terms $\Pi_X$, $Q_X$ and $W_X$ contain radial-gradient, restoring, and inertial contributions, whereas $A_{XY}$, $B_{XY}$, and $C_{XY}$ denote schematic groupings of gravitational couplings. The full component equation and background functions are given in Eqs.~(55) and (56) of Ref.~\cite{Kain2020}. Regularity is imposed at the center and $\Delta P_X=0$ at each fluid's surface. The variational reorganization used in the numerical calculation follows Ref.~\cite{Caballero2024} for which we solve Eqs. (\ref{eq:pulsation}) numerically through the Rayleigh-Ritz method improved by a finite-element discretization to extract the eigenvalues.

As proven in Ref. \cite{Caballero2024}, Eq. (\ref{eq:pulsation}) represents a  self-adjoint problem with a real spectrum of $\omega^2_n$ being ordered algebraically as $\omega_0^2\leq\omega_1^2\leq\cdots$. For stable configurations, the fundamental-mode frequency is $f_0=\omega_0/(2\pi)$. We reserve the term \emph{zero mode} for an eigenmode satisfying $\omega_n^2=0$. The lowest eigenvalue determines radial stability, i.e. $\omega_0^2>0$ is stable, $\omega_0^2=0$ is marginal, and $\omega_0^2<0$ admits an exponentially growing perturbation signaling unstable solutions. We calculate static critical curves (from Eq. \eqref{eq:jacobian}) and radial eigenfrequencies \rev{separately and compare them over their common sampled domain}. Radial stability is assigned from the sign of $\omega_0^2$, \rev{not from $J=0$ in Eq. \eqref{eq:jacobian} alone}. A vanishing excited-mode eigenfrequency inside a region with $\omega_0^2<0$ changes the unstable-mode count without restoring stability.

\subsection{Compact-Object Classification}
\label{sec:classification}

We classify our two-fluid equilibrium configurations using the total gravitational mass $M$, the OM mass $M_{\OM}$, and the radii
$R_{\OM}$ and $R_{\DM}$.  The radius $R_{\OM}$ represents the potentially
visible surface of the object. For practical purposes, a configuration is regarded as DM admixed when $F_{\DM}\geq 10^{-11}$, where $F_{\DM}=M_{\DM}/M$.  This small threshold removes numerical roundoff contamination and it should not be considered an astrophysical lower bound.

In Table~\ref{tab:primary_classes}, we define our primary objects of study: dark compact planets (DCPs), dark white dwarfs (DWDs), and dark-matter-admixed neutron stars (\rev{DANSs}). These are operational mass--radius windows, with no central-pressure condition. We stress that throughout this work, counts, rejection fractions, and medians describe the numerical sampling, not astrophysical abundances. For instance, the $0.05\Msun$ division changes labels, not the equilibrium sample. \rev{In total, our sample contains 621270 DMACO configurations with pairs of central pressures for the eight cases combining different $m_f$ and $y$. This number of objects comes from demanding the same ranges for the mass and radius axis to get clear comparisons.}

\begin{table}[!tbp]
\centering
\small
\renewcommand{\arraystretch}{1.08}
\setlength{\tabcolsep}{4pt}
\begin{ruledtabular}
\begin{tabular}{@{}p{0.24\columnwidth}p{0.68\columnwidth}@{}}
Class & Mass and Radius Conditions \\
\hline
$\DCP$ &
$M<0.05\,\Msun$ and $R_{\OM}<1000~\mathrm{km}$ \\
$\DWD$ &
$M\geq0.05\,\Msun$ and $R_{\OM}>2000~\mathrm{km}$ \\
$\DMANS$ &
$0.05\leq M/\Msun\leq3$,
$M_{\OM}\geq0.05\,\Msun$, and
$0.5\leq R_{\OM}/\mathrm{km}\leq30$ \\
Residual &
Any stable DM-admixed configuration outside the three windows above \\
\end{tabular}
\end{ruledtabular}
\caption{\label{tab:primary_classes}
Primary compact-object classes.  All configurations must contain DM
with $F_{\DM}\geq 10^{-11}$ (see main text).}
\end{table}

The subtypes in Table~\ref{tab:dcp_subtypes} are intersected with the DCPs
window, including its upper mass bound. Table~\ref{tab:structural_tags} defines two secondary objects found in our parameter-space scanning. Some radially stable DM-admixed configurations satisfy none of the three primary mass--radius windows; we call them \rev{residual configurations}. The dark-oyster-like (DOY) \cite{Biesdorf2025} selection can overlap \rev{DANS} and residual configurations, but its cuts exclude DCPs and DWDs. The extreme-halo objects are independent of the primary windows.

\begin{table}[!tbp]
\centering
\small
\renewcommand{\arraystretch}{1.08}
\begin{ruledtabular}
\begin{tabular}{@{}ll@{}}
Subtype & Total-mass interval \\
\hline
Earth-like &
$0.5<M/\Mearth<5$ \\
Jupiter-like &
$0.5<M/\Mjup<2.5$ \\
Brown-dwarf-like &
$13<M/\Mjup<80$ \\
\end{tabular}
\end{ruledtabular}
\caption{\label{tab:dcp_subtypes}
Planetary/substellar masses for DCPs.}
\end{table}

\begin{table}[!tbp]
\centering
\small
\renewcommand{\arraystretch}{1.08}
\setlength{\tabcolsep}{4pt}
\begin{ruledtabular}
\begin{tabular}{@{}p{0.30\columnwidth}p{0.62\columnwidth}@{}}
Objects & Conditions \\
\hline
$\DOY$ &
$0.1<M/\Msun<3$,
$R_{\OM}<25~\mathrm{km}$, and
$R_{\DM}>4R_{\OM}$ \\
Dark extreme halo &
$R_{\DM}/R_{\OM}>100$ \\
\end{tabular}
\end{ruledtabular}
\caption{\label{tab:structural_tags}
Secondary objects defined by masses and radii.}
\end{table}

\begin{figure*}[!t]
\centering
\articlegraphic[width=0.96\textwidth]{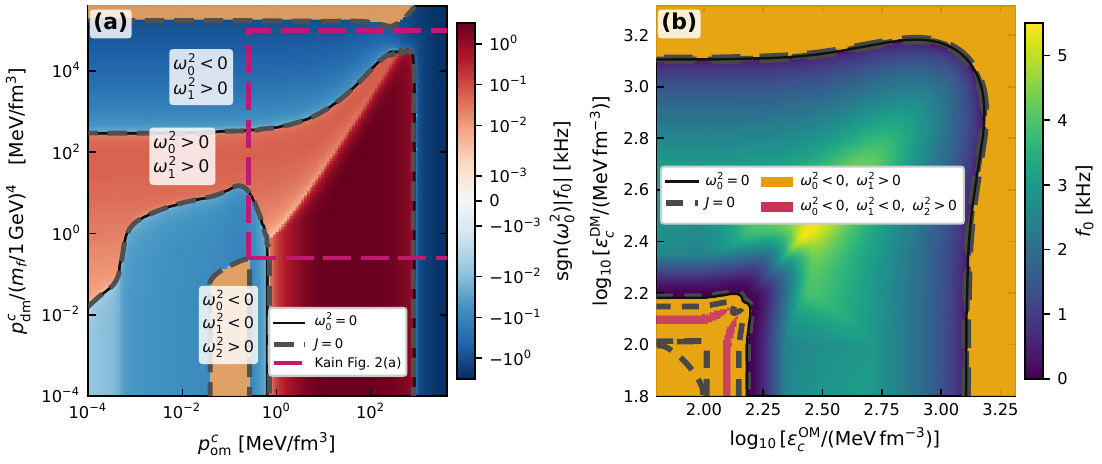}
\caption{Reference calculations to test our two-fluid pulsation code. {\it Left panel:}  (a) Results for the pressure space for the SLy EoS admixed with fermionic DM ($m_f=0.2~\mathrm{GeV}$ and $y=0$). The accompanying frequency colormap gives $\operatorname{sgn}(\omega_0^2)|f_0|$, the light-orange overlay
marks two negative modes, while the black curves refer to $\omega_0^2=0$, the dashed gray curves are $J=0$ for the turning-point criterion, and the magenta dashed rectangle reproduces the pressure window of Fig.~2(a) in Ref.~\cite{Kain2021}. {\it Right panel:} (b) Similar results as those of Ref. \cite{KumarSotani2025} for QMC-RMF4 with mirror DM. The stable domain is colored by $f_0$; orange and red bands mark one and two negative modes, respectively. The black and dashed-gray curves have the same
meaning as in panel (a).}
\label{fig:external_benchmark}
\end{figure*}

Only those DMACOs whose mass is mainly carried by DM are described as DM dominated. For each named class, we first identify candidates from equilibrium masses and radii, \rev{irrespective of their radial stability}, and then retain those with $\omega_0^2>0$. \rev{A class contour can therefore extend into an unstable region}, them implying that those configurations might meet the structural definition but also could fail the radial-oscillation test.

\section{Numerical results: Two-fluid radially stable DMACOs}
\label{sec:results}

Before computing the two-fluid fundamental-mode eigenfrequencies for EoS-I admixed with DM, we tested our pulsation code against the reference models of Refs.~\cite{Kain2021,KumarSotani2025}. Figure~\ref{fig:external_benchmark}(a) shows the central-pressure plane for the SLy EoS~\cite{DouchinHaensel2001} and free fermionic DM with $m_f=0.2~\mathrm{GeV}$. The dashed magenta rectangle marks the pressure window of Fig.~2(a) of Ref.~\cite{Kain2021}, with the same rescaling of the dark-pressure axis. We recover the high-density \rev{DANS} stability boundary and resolve a small unstable pocket at slightly lower central pressures. Extending the scan to lower pressures reveals an additional stable domain, shown in red, where $\omega_0^2>0$. The blue region has \rev{one negative squared radial eigenfrequency}; \rev{the orange region has two}. Both are radially unstable, and the orange region satisfies \rev{$\omega_0^2\leq\omega_1^2<0$}. The gray $J=0$ curves therefore include both the fundamental-mode boundary and an internal curve on which the first overtone crosses zero. Crossing the latter changes the number of unstable modes without restoring stability.

Figure~\ref{fig:external_benchmark}(b) provides a second test using QMC-RMF4 and mirror DM, following Ref.~\cite{KumarSotani2025}. The calculation recovers the reported stability domains and fundamental-frequency scale. At low central pressures, resolving $J=0$ reveals additional critical curves; the pulsation spectrum identifies whether the corresponding zero belongs to the fundamental mode or to an overtone. Thus, a static critical curve is not necessarily a boundary between stable and unstable configurations.

The main implication of Fig.~\ref{fig:external_benchmark} is that $J=0$ and the radial spectrum provide complementary information. Under the assumptions of Ref.~\cite{Chen2026}, rank loss identifies a physical zero mode, but it does not label that mode. Our frequency maps identify the fundamental-mode boundary and distinguish it from overtone zeros inside an already unstable domain. These additional curves do not invalidate the generalized turning-point criterion; they show why mode identification matters when classifying low-pressure compact objects. The close approach of critical contours should not, by itself, be interpreted as an avoided crossing of radial modes~\cite{Gondek:1999ad}: establishing that behavior requires tracking the eigenfrequencies and eigenfunctions, not merely their zero contours.

We also tested the static eigenvalue diagnostic of Hippert et al.~\cite{Hippert2023}, based on $A_{ij}=\partial N_i/\partial\epsilon_j^c$. In selected very-low-pressure SLy configurations with free fermionic DM of mass $0.2~\mathrm{GeV}$, \rev{both eigenvalues of $A$ are positive despite $\omega^2_0<0$ and $\omega^2_1<0$}. We confirmed the latter with two independent numerical approaches: \rev{a finite-element eigenvalue calculation and shooting integration of the coupled perturbation equations}. Thus, \rev{local positivity of $A$ alone does not establish stability in these tested cases}; this does not contradict the necessary condition of Ref.~\cite{Hippert2023} for DANSs.

\begin{figure*}[p]
\centering
\articlegraphic[width=\textwidth,height=0.90\textheight,keepaspectratio]{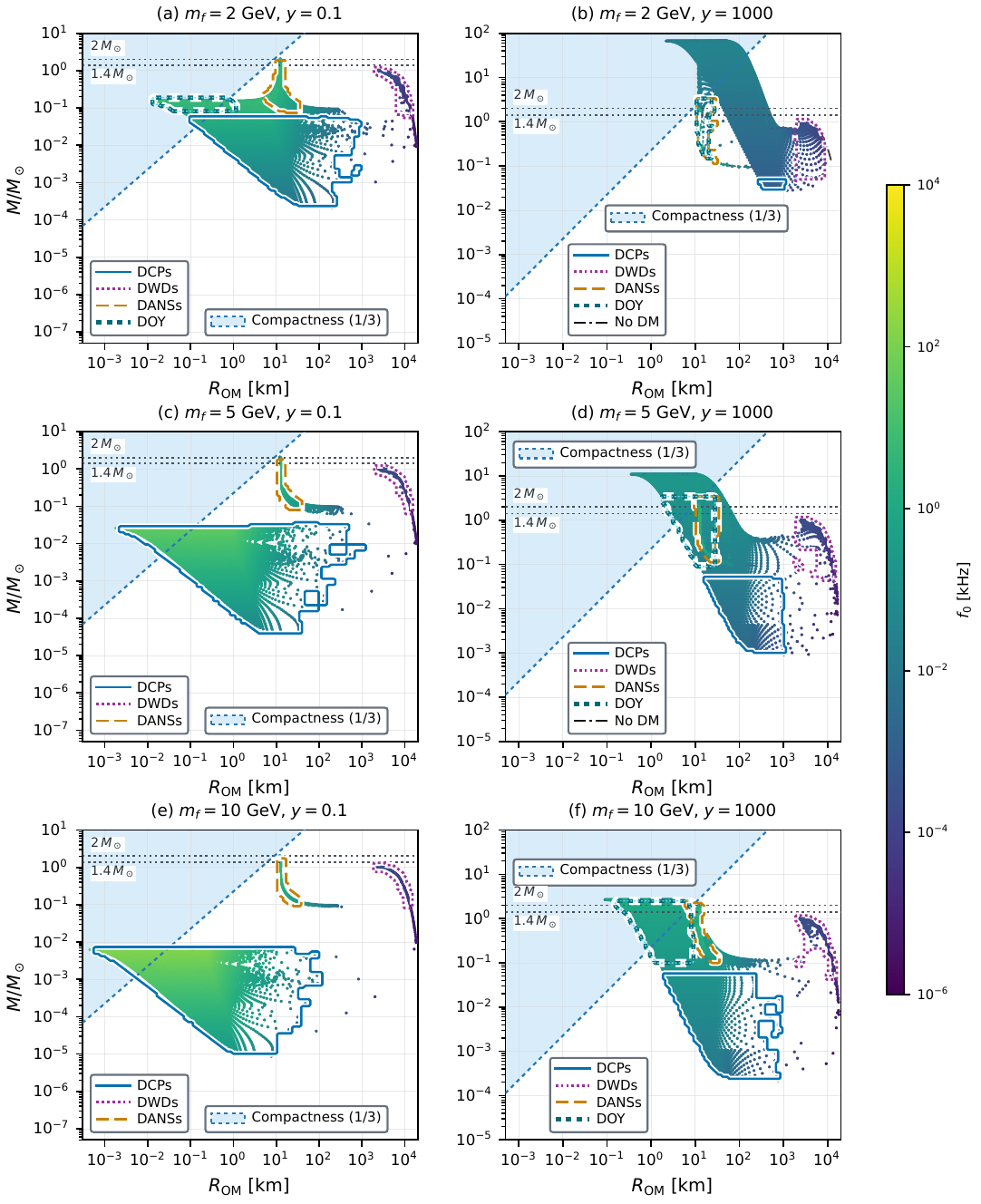}
\caption{\label{fig:mr_frequency}
\rev{Sampled radially} stable DMACOs \rev{for which} DM has $m_f=2$, 5, and
$10~\mathrm{GeV}$ (rows) and $y=0.1$, $10^3$ (columns), colored by the
fundamental radial frequency $f_0$ on the right.  Solid, dotted, and dashed enclosures mark DCP, DWD, and \rev{DANS} objects, respectively. The DOY contour overlaps with other enclosures. The dash-dotted black curves indicate the one-fluid radially stable NS and white-dwarf branches, although they are mostly obscured by the corresponding DM-admixed configurations. Axis limits are \rev{common across fermion masses at fixed $y$} and need not coincide between the two strengths. Dotted horizontal references are \rev{labelled directly at $1.4\Msun$ and $2\Msun$}. The light-blue regions mark $M/R_{\OM}\geq1/3$, \rev{a compactness reference rather than a determination of a photon sphere inside an extended halo}.}
\end{figure*}

\begin{figure*}[p]
\centering
\articlegraphic[width=\textwidth,height=0.90\textheight,keepaspectratio]{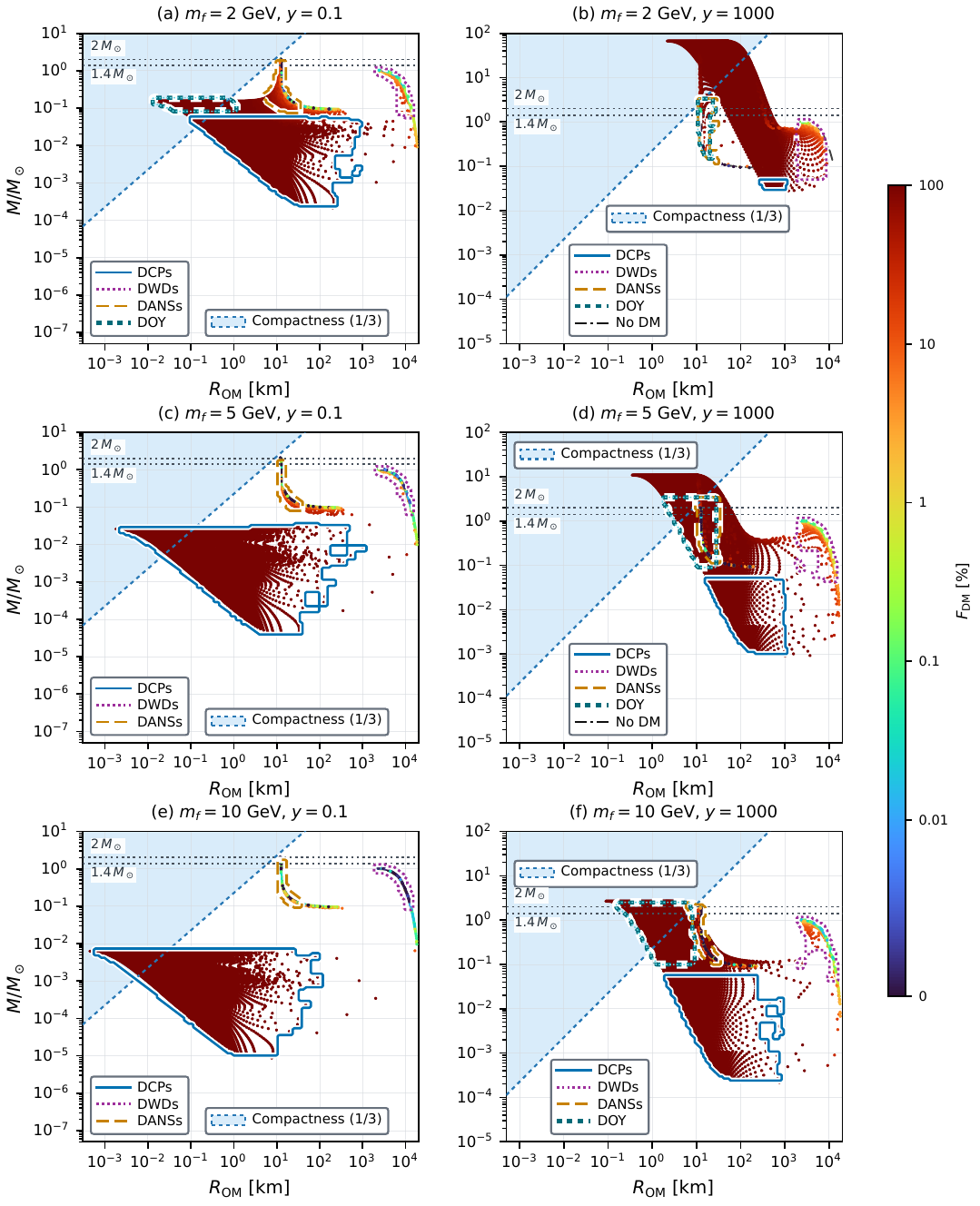}
\caption{\label{fig:mr_fraction}
Same \rev{sampled} stable \rev{configurations and common axes} as Fig.~\ref{fig:mr_frequency}, now colored by $100F_{\DM}$ \rev{in percent}. \rev{Class} envelopes and \rev{the compactness} reference have the same definitions and \rev{limitations}.}
\end{figure*}

\begin{figure*}[!t]
\centering
\articlegraphic[width=1\textwidth]{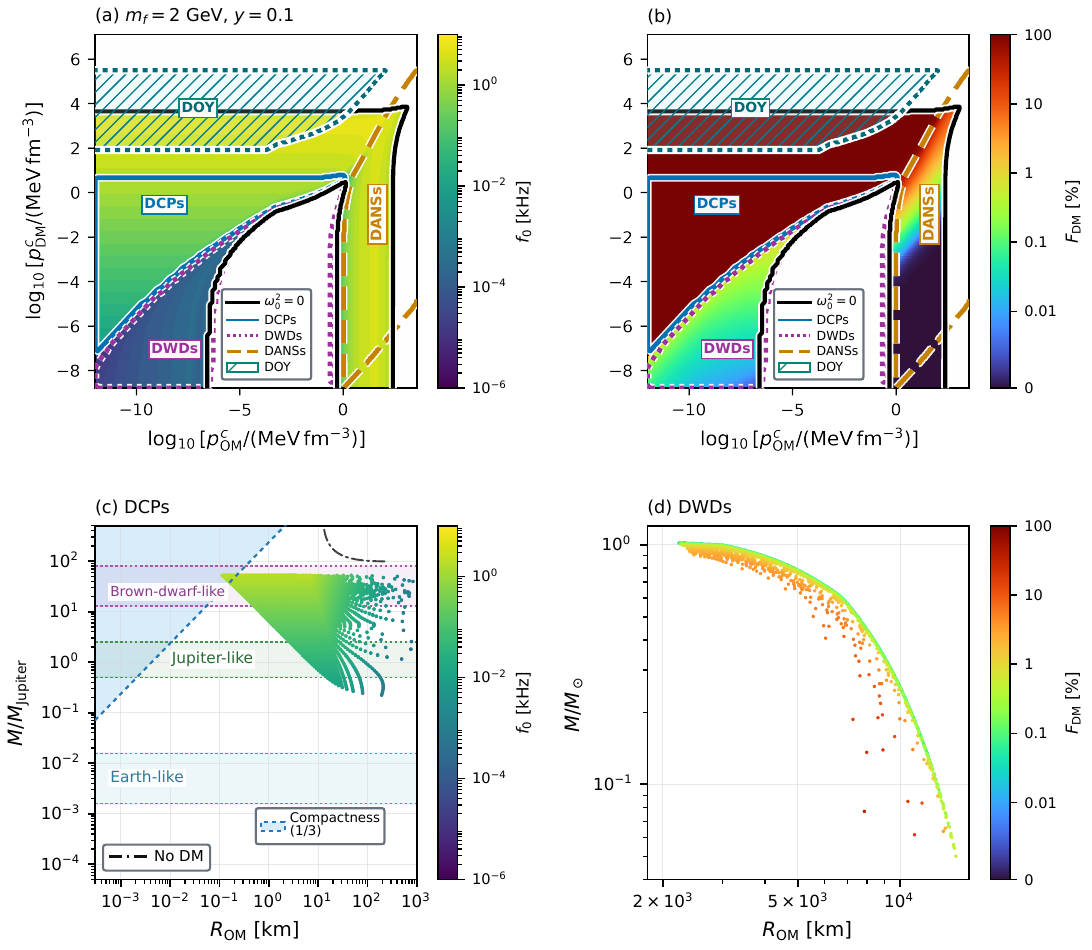}
\caption{\label{fig:weak2}
Various results for the OM EoS-I admixed with DM having  $m_f=2~\mathrm{GeV}$ and $y=0.1$. Panels (a) and (b)  show the physical central-pressure plane colored by $f_0$ and the DM mass percentage $100F_{\DM}$, respectively. The black contour is $\omega_0^2=0$. Colored class contours \rev{follow} equilibrium \rev{mass and radius criteria and can extend outside the radially stable set}. \rev{Blank space alone must not be interpreted as a stability classification.} DOY hatching denotes the structural criterion in Table~\ref{tab:structural_tags}. Panels (c) and (d) show dynamically stable DCP and DWD configurations, colored by $f_0$ and $100F_{\DM}$, respectively. Pressure and DCP axes are \rev{common across masses at fixed $y$}. \rev{DCP panel masses are in Jupiter masses.} In the DWD panel, \rev{dark-purple points are low-DM-fraction configurations, not an additional class}; all selected points are retained, with \rev{smaller markers and low fractions drawn first}. The one-fluid and compactness references are defined in Fig.~\ref{fig:mr_frequency}.}
\end{figure*}

\begin{figure*}[!t]
\centering
\articlegraphic[width=1\textwidth]{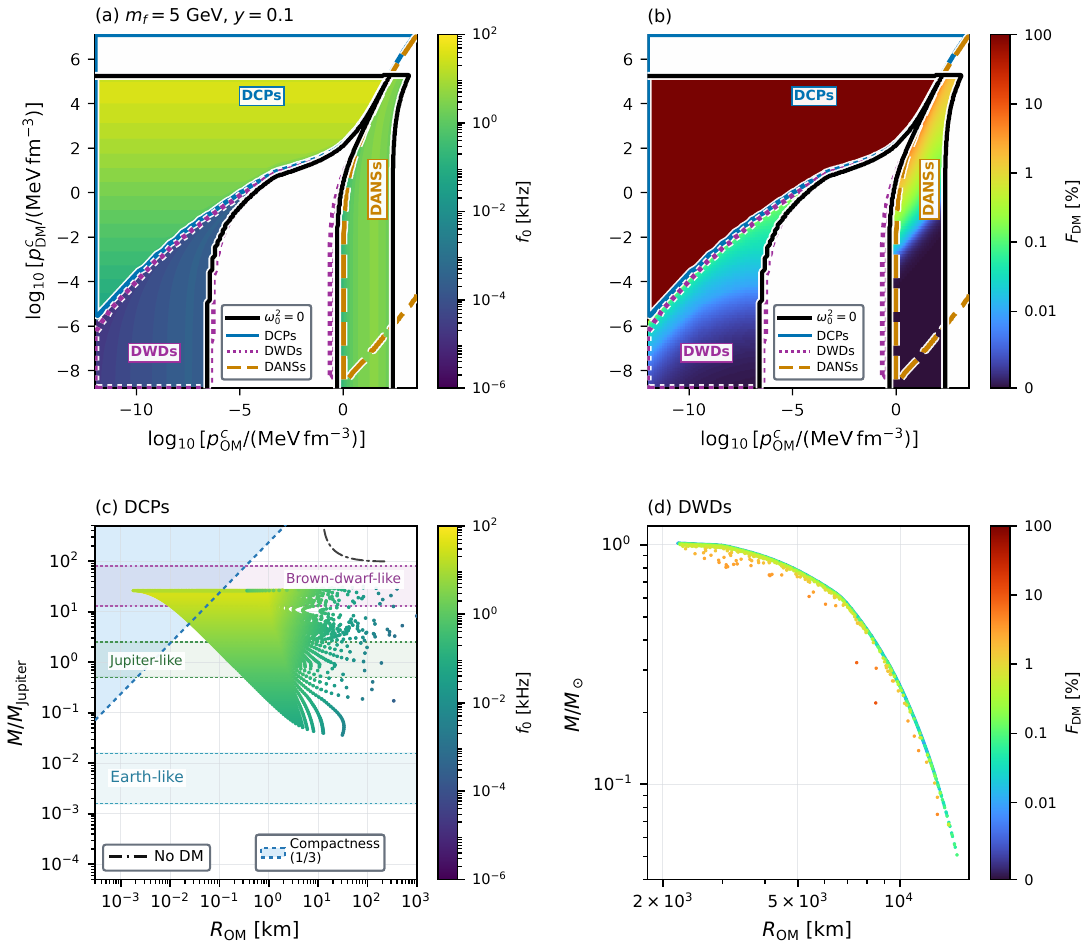}
\caption{\label{fig:weak5}
Same as Fig.~\ref{fig:weak2} but for $m_f=5~\mathrm{GeV}$ and $y=0.1$. In this case, no stable DOY configurations are found.}
\end{figure*}

\begin{figure*}[!t]
\centering
\articlegraphic[width=1\textwidth]{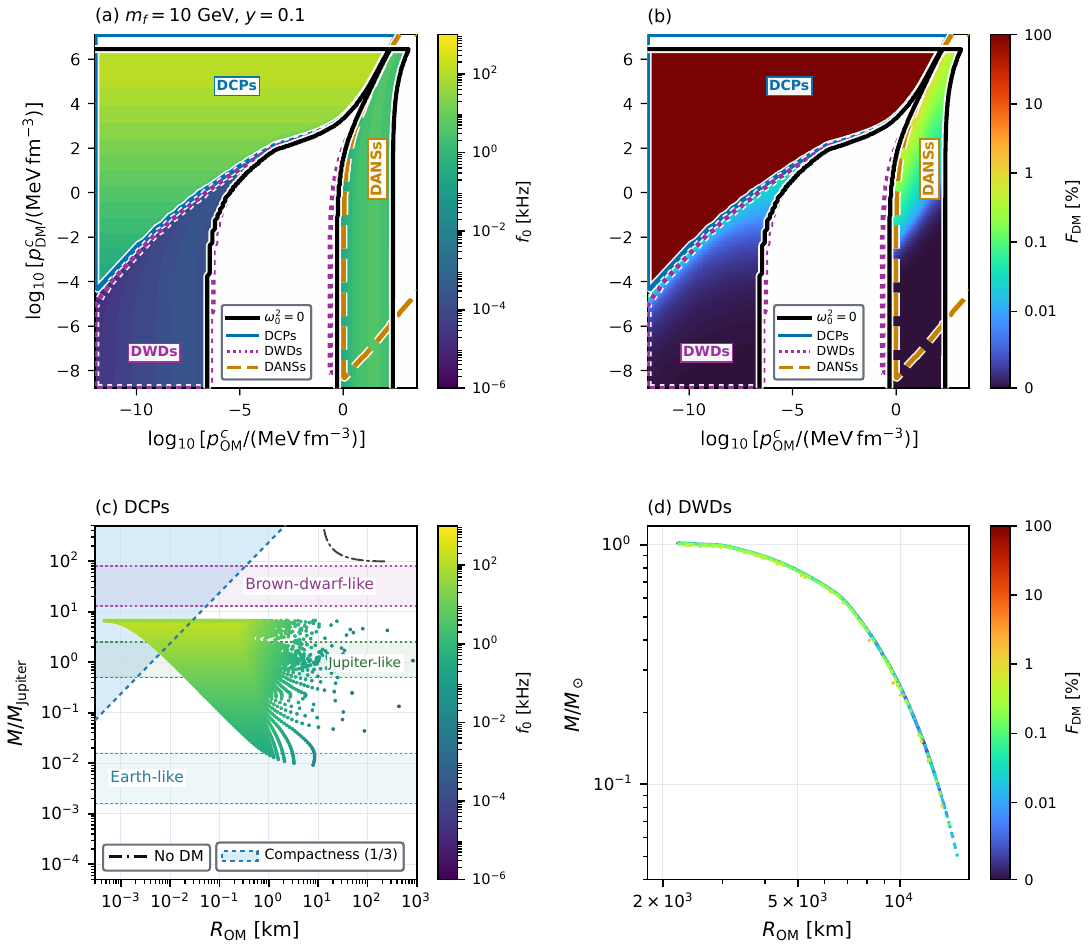}
\caption{\label{fig:weak10}
Same as Fig.~\ref{fig:weak2} but for $m_f=10~\mathrm{GeV}$ and $y=0.1$. Some stable Earth-like planets start to populate.}
\end{figure*}

\subsection{Two-Fluid Mass--Radius Structure}
\label{sec:mr_overview}

Now we solve the two-fluid TOV and pulsation equations for the OM EoS-I and DM EoS for several fermion masses and self-couplings. The numerical integration also includes central OM pressures corresponding to radially unstable objects in the one-fluid limit. This allows us to scan the pressure space from extremely low to high central pressures for both fluids without any additional constraint that might exclude objects that become stable when admixed with DM. Then, we show in Fig.~\ref{fig:mr_frequency} the two-fluid radially stable mass-radius ($M-R$) relations, where one can see how the DM microphysics reorganizes the one-fluid stable branches (NSs and white dwarfs) and naturally introduces  the appearance of new DM admixed objects. Notice that in these plots (and those of Fig.~\ref{fig:mr_fraction}) we also added horizontal \rev{reference} lines \rev{at $2\Msun$ and $1.4\Msun$, labelled directly on the plots. Neither line is imposed as a universal maximum mass for the mixed configurations}. Moreover, such $M$--$R$ diagrams commonly show reference bounds on the compactness (defined as $M/R_{\OM}$) such as the Schwarzschild radius ($1/2$), Buchdahl (4/9) and causality (1/2.9) limits \cite{Schaffner:2020}. \rev{We also show the compactness reference $R_{\OM}=3M$~\cite{Cardoso:2019rvt,Pitz:2024xvh} typical of a photon sphere in one-fluid stars. When a DM halo extends beyond the OM surface}, \rev{this 1/3 bound does not by itself locate a photon sphere inside the material halo; that requires the corresponding metric}.

In the weakly repulsive case ($y=0.1$), the fermionic degeneracy pressure sets the new admixed mass and radius scales. One can see that increasing $m_f$ shifts the DM-supported branches toward smaller masses and radii, while the DWD sequences remain close to the corresponding OM branch. Strong self-repulsion ($y=10^3$) stiffens the DM EoS and allows the admixed stars to support much more massive and spatially extended objects around a compact OM core. In particular, for $m_f=2$ and $5~\mathrm{GeV}$, the strongly repulsive sequences extend beyond $3\Msun$, outside our primary \rev{DANS} mass window. As expected, these high-mass stable configurations follow the $m_f^{-2}$ scaling of self-interacting fermion
stars found in Ref. \cite{Narain2006}. Besides, the OM mass becomes negligible near the high-mass end of these two-fluid sequences, approaching the pure-DM limit. \rev{These DM-dominated configurations} contain compact OM cores \rev{surrounded by extended dark halos}, as discussed in Sec.~\ref{sec:strong}. These structural trends are consistent with Refs.~\cite{Tolos2015,Dengler2022}; here we \rev{additionally assess stability with the coupled radial-pulsation problem}.

The fundamental-mode frequency colors in Fig.~\ref{fig:mr_frequency}, together with the composition map in Fig.~\ref{fig:mr_fraction}, reveal a separation of characteristic oscillation scales between the stable DMACO families. DCPs are generally DM dominated, and their frequency scale changes strongly with $m_f$ and $y$. By contrast, DWDs have modest median DM fractions and typical frequencies of order $2\times10^{-4}~\mathrm{kHz}$, while OM-dominated \rev{DANSs} retain frequencies of a few kHz. This contrast is consistent with the DM fluid setting the principal structural scale of DCPs, while the OM fluid largely sets the characteristic DWD and \rev{DANS} scales. \rev{Small DWD frequencies away from that boundary can instead reflect their extended, low-density structure.} The DOY selection is geometric, based on masses and radii: \rev{a large dark-to-ordinary radius ratio can occur with either a moderate or a dominant DM mass fraction}.

The horizontal axes show $R_{\OM}$, but the dark radii $R_{\DM}$ are also available for each configuration, \rev{although not displayed in the plots}. Among sampled stable configurations with $M/R_{\OM}\geq1/3$, the median $R_{\rm DM}$ for $m_f=2$, 5, and $10~\mathrm{GeV}$ are \rev{$2.65$, $0.424$, $0.106$} $\mathrm{km}$ at $y=0.1$, with interquartile intervals \rev{$2.3$--$3.32$, $0.369$--$0.531$, $0.0922$--$0.133$} $\mathrm{km}$. At $y=10^3$, the corresponding medians are \rev{$586$, $91.3$, $22.7$} $\mathrm{km}$ and the intervals are \rev{$520$--$650$, $81$--$101$, $20.3$--$25.3$} $\mathrm{km}$. The unchanged $100~\mathrm{GeV}$ sample has no member satisfying this compactness condition at $y=0.1$. At $y=10^3$, its median is $0.230~\mathrm{km}$ and its interquartile interval is $0.205$--$0.253~\mathrm{km}$. The OM surface can lie inside $R_{\rm OM}<3M$, while the DM halo extends beyond it.

\begin{figure*}[!t]
\centering
\articlegraphic[width=1\textwidth]{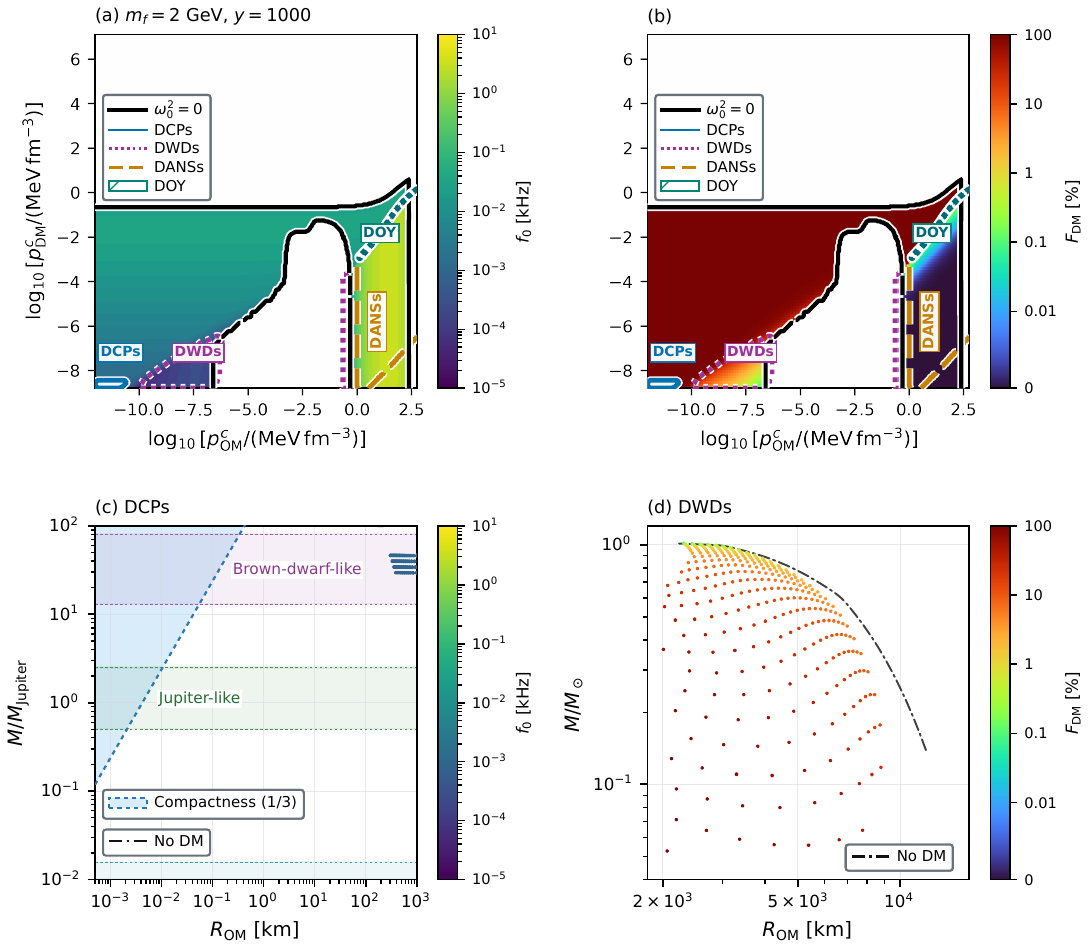}
\caption{\label{fig:strong2}
Various results for OM admixed with strongly-repulsive ($y=10^3$) DM for $m_f=2~\mathrm{GeV}$ arranged as in Fig.~\ref{fig:weak2}.}
\end{figure*}

\begin{figure*}[!t]
\centering
\articlegraphic[width=1\textwidth]{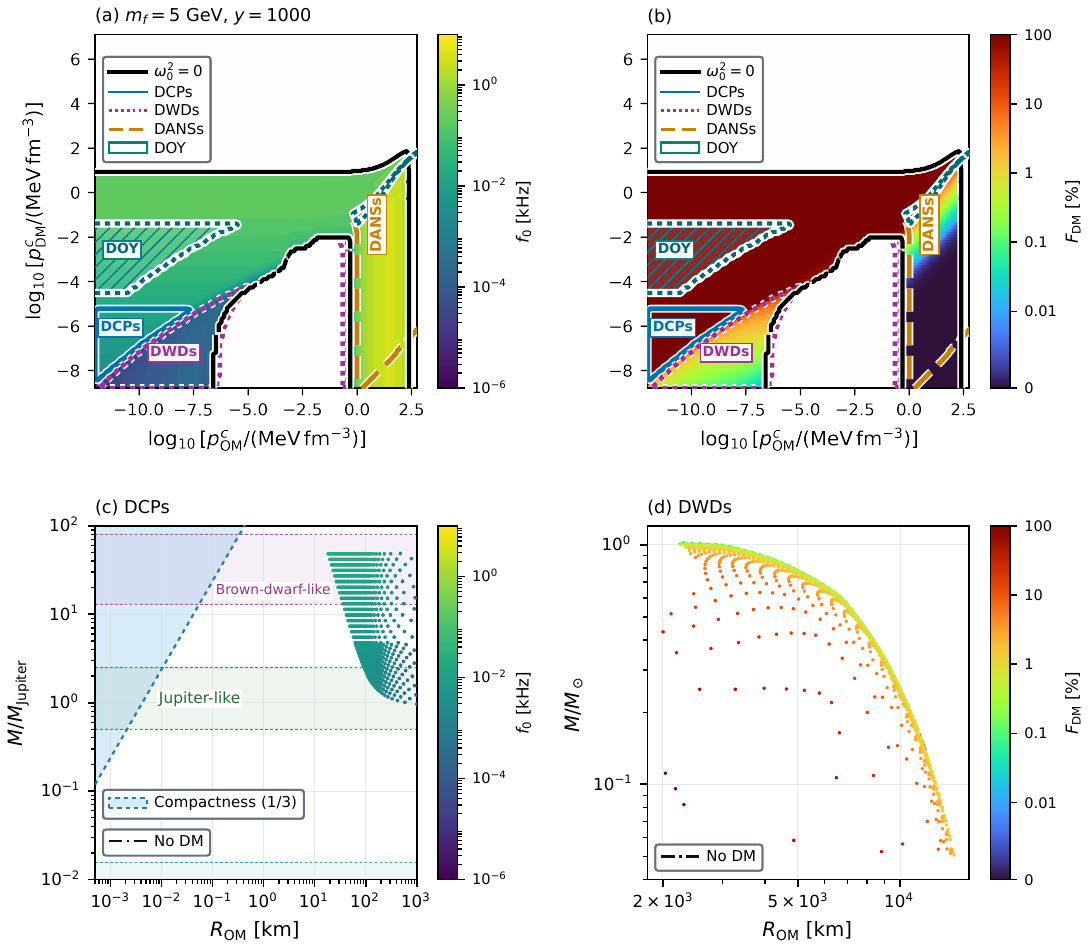}
\caption{\label{fig:strong5}
Same as Fig.~\ref{fig:strong2} but for $m_f=5~\mathrm{GeV}$.  Unlike the previous case, DOYs are now significantly present.}
\end{figure*}

\begin{figure*}[!t]
\centering
\articlegraphic[width=1\textwidth]{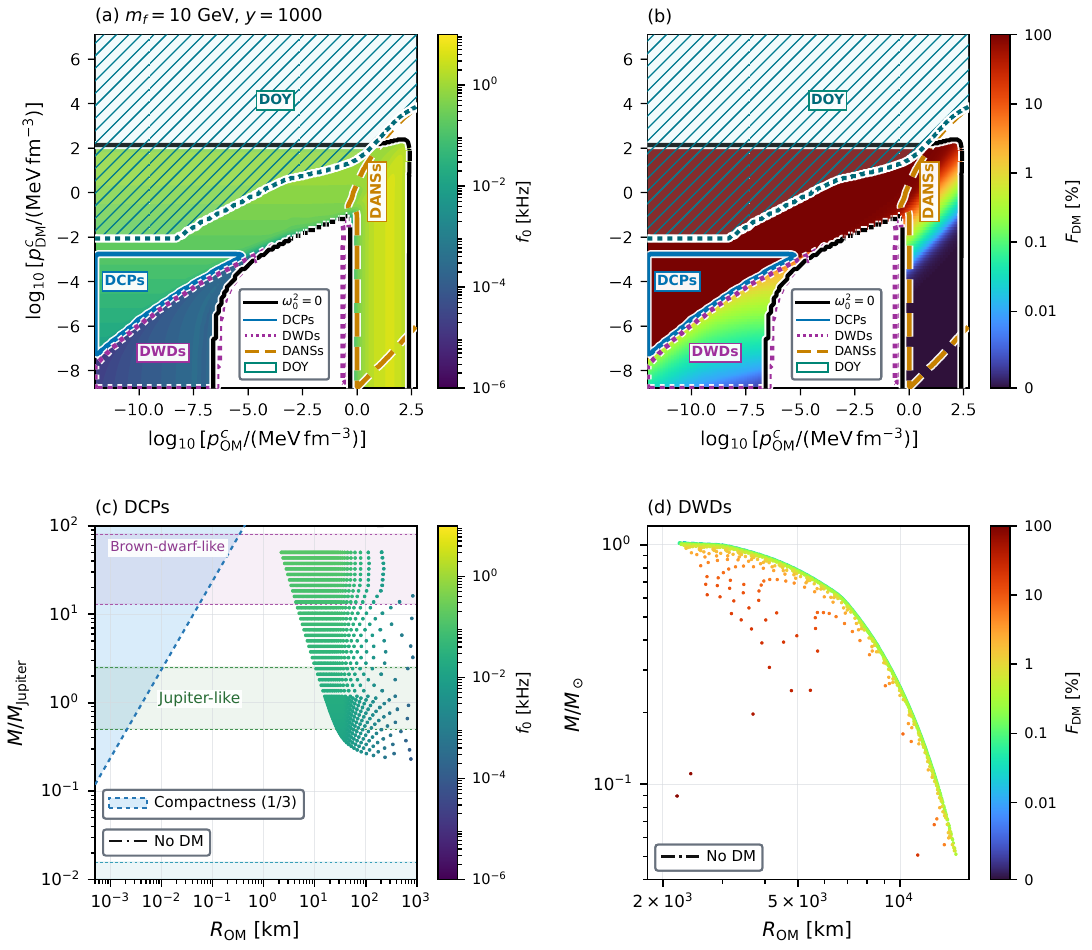}
\caption{\label{fig:strong10}
Same as Fig.~\ref{fig:strong2} but for $m_f=10~\mathrm{GeV}$. In this case the DOY region covers a broad part of the stable, highly DM-dominated halo branch.}
\end{figure*}

\subsection{Weakly-Repulsive DMACOs}
\label{sec:weak}

Figures~\ref{fig:weak2}--\ref{fig:weak10} show the sampled stable central-pressure domains for $y=0.1$, colored by $f_0$ and $100F_{\DM}$, together with DCP and DWD mass--radius projections. Class contours use only the mass--radius definitions, independently of the stability test. Within our sample, the DWD rejection fractions are \rev{$33.31\%$, $24.63\%$, and $21.93\%$} for $m_f=2$, 5, and $10~\mathrm{GeV}$; the corresponding DANS fractions are \rev{$30.90\%$, $30.03\%$, and $28.00\%$} (Table~\ref{tab:weak_dynamic_rejection}). DCP rejection is \rev{zero, $36.26\%$, and $15.97\%$}, respectively. Only the $2~\mathrm{GeV}$ case contains sampled DOY candidates, of which \rev{$62.76\%$} are rejected.

\begin{table*}[!tbp]
\centering
\small
\setlength{\tabcolsep}{3pt}
\renewcommand{\arraystretch}{1.07}
\begin{ruledtabular}
\begin{tabular}{@{}lccc@{}}
$m_f$ [GeV] & 2 & 5 & 10 \\
\hline
\multicolumn{4}{c}{Rejected configurations [\%]} \\
\hline
DCP & \rev{0.00} & \rev{36.26} & \rev{15.97} \\
DWD & \rev{33.31} & \rev{24.63} & \rev{21.93} \\
\rev{DANS} & \rev{30.90} & \rev{30.03} & \rev{28.00} \\
DOY & \rev{62.76} & \rev{--} & \rev{--} \\
\hline
$M_{\rm max}/\Msun$
      & 1.987  & 1.987  & 1.987  \\
$f_0(M_{\rm max})$ [kHz]
      & 0.1388 & 0.1388 & 0.1388 \\
\end{tabular}
\end{ruledtabular}
\caption{\label{tab:weak_dynamic_rejection}
Percentage of weakly repulsive ($y=0.1$) DMACO candidates rejected by the radial stability analysis, \rev{$100N_{\rm unstable}/(N_{\rm stable}+N_{\rm unstable})$}, among candidates with an assigned radial-stability sign. \rev{Unclassified candidates are excluded from this denominator.} Percentages depend on the sampled grid and domain, \rev{not on an astrophysical population}. Dashes mean that no DOY candidates occur in the sampled domain. The \rev{last} rows give the largest sampled stable mass \rev{and the frequency at that same configuration}, \rev{not at an exactly located zero mode}. \rev{This convention also applies to Table~\ref{tab:strong_dynamic_rejection}}.}
\end{table*}

For $m_f=2$, 5, and $10~\mathrm{GeV}$, the median DCP frequencies are \rev{$0.642$, $16.1$, $64.5$} $\mathrm{kHz}$, respectively. The corresponding DWD medians are \rev{$0.00017$, $0.000148$, $0.000138$} $\mathrm{kHz}$, and the DANS medians are \rev{$2.72$, $2.6$, $2.61$} $\mathrm{kHz}$. Increasing the fermion mass therefore changes the DCP frequency scale much more strongly than that of DWDs or DANSs.

By analyzing in more detail the two-fluid TOV solutions for $m_f=2~\mathrm{GeV}$, we realized that in most cases, small OM cores are surrounded by extended dark halos. Besides, stable \rev{configurations} outside the primary windows in Tables \ref{tab:primary_classes}--\ref{tab:structural_tags} are strongly DM dominated with a median of around $M=0.116\Msun$, $R_{\OM}=0.910~\mathrm{km}$, $R_{\DM}=3.19~\mathrm{km}$, and $F_{\DM}=99.9992\%$. Their subsolar masses
distinguish them from the more massive DM-dominated family found for
strong repulsion to be analyzed later. Interestingly, just for this $m_f$ we were able to find stable DOYs.

On the other hand, our two-fluid TOV results for $m_f=5$ and $10~\mathrm{GeV}$ indicate that the DM fluid becomes much more compact and no stable configuration satisfies the DOY criterion. Furthermore, radially stable configurations outside the primary windows are now mainly OM-dominated having intermediate OM radius objects with a median of around $M\simeq0.10\Msun$ and $R_{\OM}\simeq54~\mathrm{km}$, rather than a continuation of the extended halo family for $m_f=2~\mathrm{GeV}$. 

Remarkably, DWDs become more compact as $m_f$ increases, approaching the masses and radii of their one-fluid counterparts. For $m_f=10~\mathrm{GeV}$, their mass--radius sequences closely resemble those of ordinary white dwarfs, making them difficult to distinguish through these properties alone in future observations.

\subsection{Strongly-Repulsive DMACOs}
\label{sec:strong}

Strong repulsive self-interactions with $y=10^3$ yield a stiffer DM EoS and support massive, extended dark halos around compact OM cores. The largest sampled stable mass reaches $66.79\Msun$ for $m_f=2~\mathrm{GeV}$, with lower maxima of $10.69\Msun$ and $2.672\Msun$ for 5 and $10~\mathrm{GeV}$, respectively (Table~\ref{tab:strong_dynamic_rejection}). Their approximate $m_f^{-2}$ scaling and negligible OM mass at the high-mass end connect these configurations to the self-interacting fermion-star limit~\cite{Narain2006}. They are therefore not ordinary NSs whose mass has been raised by a small dark admixture: their dominant mass and structural support reside in the dark fluid. \rev{For fixed DM parameters}, \rev{DCPs, DWDs}, and \rev{DANSs are selected from the same two-fluid family by different mass--radius} windows.

Interestingly, every classified DCP candidate in the strong-interaction sample is radially stable for $m_f=2$, 5, and $10~\mathrm{GeV}$. DWD rejection is \rev{$50.37\%$, $28.95\%$, and $24.22\%$}, respectively. The corresponding DANS rejection fractions are \rev{$21.73\%$, $19.31\%$, and $23.47\%$}, while DOY rejection is \rev{$14.93\%$, $2.60\%$, and $81.47\%$} (see Table~\ref{tab:strong_dynamic_rejection}). In Tables~\ref{tab:weak_dynamic_rejection} and \ref{tab:strong_dynamic_rejection}, $M_{\rm max}$ is the largest sampled stable mass, and \rev{the frequency belongs to that same grid point, not to an exactly located zero mode}.

For strong repulsion and $m_f=2$, 5, and $10~\mathrm{GeV}$, the median DCP frequencies are \rev{$0.00105$, $0.00705$, $0.0539$} $\mathrm{kHz}$, respectively. The corresponding DWD medians are \rev{$0.000216$, $0.000146$, $0.00013$} $\mathrm{kHz}$, while DANS medians are \rev{$2.32$, $2.26$, $2.2$} $\mathrm{kHz}$. Strong self-repulsion expands the DM fluid and lowers the characteristic DCP frequencies relative to weak repulsion. The DWD and DANS frequency scales change less, consistent with their OM-dominated median structure. The mass--radius relations are shown in Figs.~\ref{fig:strong2}--\ref{fig:strong10}.

For $m_f=2~\mathrm{GeV}$ and $y=10^3$, the residual sample is \rev{DM dominated at the median and extends beyond the primary DANS mass window}. Table~\ref{tab:other_properties} lists its component medians. For the separately selected extreme-halo subset, the median masses and radii are \rev{$(M/\Msun,R_{\OM}/\mathrm{km},R_{\DM}/\mathrm{km})=(64.4, 4.21, 519)$}. \rev{These separately computed medians does not describe one configuration.}

For $m_f=5~\mathrm{GeV}$ and $y=10^3$, the residual sample is \rev{DM dominated at the median and extends beyond the primary DANS mass window}. Table~\ref{tab:other_properties} lists its component medians. For the separately selected extreme-halo subset, the median mass and radii are \rev{$(M/\Msun,R_{\OM}/\mathrm{km},R_{\DM}/\mathrm{km})=(10.3, 0.673, 83.1)$}. \rev{These separately computed medians need not describe one configuration.}

For $m_f=10~\mathrm{GeV}$ and $y=10^3$, residual stable configurations can have total masses in the NS range while \rev{their OM contribution fails the DANS mass requirement}. Their medians indicate a compact OM core in a DM halo (Table~\ref{tab:other_properties}). The extreme-halo subset has separately computed medians \rev{$(M/\Msun,R_{\OM}/\mathrm{km},R_{\DM}/\mathrm{km})=(2.58, 0.168, 20.8)$}.

\begin{table*}[!tbp]
\centering
\small
\setlength{\tabcolsep}{3pt}
\renewcommand{\arraystretch}{1.07}
\begin{ruledtabular}
\begin{tabular}{@{}lccc@{}}
$m_f$ [GeV] & 2 & 5 & 10 \\
\hline
\multicolumn{4}{c}{Rejected configurations [\%]} \\
\hline
DCP & \rev{0.00} & \rev{0.00} & \rev{0.00} \\
DWD & \rev{50.37} & \rev{28.95} & \rev{24.22} \\
\rev{DANS} & \rev{21.73} & \rev{19.31} & \rev{23.47} \\
DOY & \rev{14.93} & \rev{2.60} & \rev{81.47} \\
\hline
$M_{\rm max}/\Msun$
      & 66.79   & 10.69  & 2.672 \\
$f_0(M_{\rm max})$ [kHz]
      & 0.00474 & 0.0296 & 0.118 \\
\end{tabular}
\end{ruledtabular}
\caption{\label{tab:strong_dynamic_rejection}
Same as Table\rev{~}\ref{tab:weak_dynamic_rejection} for $y=10^3$\rev{, including the finite-grid convention for $M_{\rm max}$ and its nonzero frequency}.}
\end{table*}

\subsubsection*{Dark-matter-dominated configurations}
\label{sec:dm_dominated}

Many strongly repulsive residuals are DM dominated, whereas \rev{the weakly repulsive residuals at 5 and $10~\mathrm{GeV}$ are OM dominated at the median} (Table~\ref{tab:other_properties}). Strongly repulsive DCPs are nearly pure DM in the sampled cases but meet the low-mass selection $M<0.05\Msun$ and $R_{\OM}<1000~\mathrm{km}$. An extreme halo satisfies $R_{\DM}/R_{\OM}>100$. The DOY selection instead requires $0.1<M/\Msun<3$, $R_{\OM}<25~\mathrm{km}$, and $R_{\DM}/R_{\OM}>4$. \rev{Some configurations satisfy both selections, but neither replaces the other.} The extreme DM halo medians quoted above refer to \rev{that subset, not to all residuals in Table~\ref{tab:other_properties}}. \rev{Maximum total masses and maximum $R_{\rm DM}$ can occur for different configurations}.

\begin{table*}[!tbp]
\small
\setlength{\tabcolsep}{3.5pt}
\centering
\begin{ruledtabular}
\begin{tabular}{cccccccc}
$y$ & $m_f$ [GeV] & $M/\Msun$ & $M_{\OM}/\Msun$ &
$R_{\OM}$ [km] & $R_{\DM}$ [km] & $R_{\DM}/R_{\OM}$ & $100F_{\DM}$ \\
\hline
$0.1$ & 2 & \rev{$0.11632$} & \rev{$7.668\times10^{-7}$} & \rev{$0.91$} & \rev{$3.19$} & \rev{$3.5565$} & \rev{$99.999$} \\
$0.1$ & 5 & \rev{$0.10106$} & \rev{$0.1006$} & \rev{$53.65$} & \rev{$0.6981$} & \rev{$0.0080612$} & \rev{$0.1034$} \\
$0.1$ & 10 & \rev{$0.10486$} & \rev{$0.1043$} & \rev{$53.61$} & \rev{$0.2255$} & \rev{$0.002824$} & \rev{$0.015344$} \\
$10^3$ & 2 & \rev{$44.401$} & \rev{$0.05188$} & \rev{$45.74$} & \rev{$614.4$} & \rev{$12.129$} & \rev{$99.888$} \\
$10^3$ & 5 & \rev{$6.3352$} & \rev{$0.003207$} & \rev{$14.2$} & \rev{$99.06$} & \rev{$6.2971$} & \rev{$99.935$} \\
$10^3$ & 10 & \rev{$1.5828$} & \rev{$2.886\times10^{-5}$} & \rev{$4.854$} & \rev{$25.27$} & \rev{$4.9033$} & \rev{$99.998$} \\
\end{tabular}
\end{ruledtabular}
\caption{\label{tab:other_properties}
For each fixed $(m_f,y)$, the table summarizes radially stable DM-admixed configurations outside all three primary windows of Table~\ref{tab:primary_classes}. Each column gives a separately calculated median: \rev{the median mass and median radius does not belong to the same configuration}. The ratio column is \rev{the median of individual $R_{\DM}/R_{\OM}$ values, not the ratio of the two median radii}. The last column gives $100F_{\DM}$ in percent.}
\end{table*}

\begin{table}[!tbp]
\setlength{\tabcolsep}{2.2pt}
\begin{ruledtabular}
\begin{tabular}{ccrrc}
$y$ & Class & $N_{\rm cand}$ & $N_{\rm rej}$ & Rejected [\%] \\
\hline
\multirow{3}{*}{$0.1$}
 & DCP   & $26\,497$ & $19\,377$ & $73.13$ \\
 & DWD   & $3\,035$  & $965$     & $31.80$ \\
 & \rev{DANS} & $20\,764$ & $5\,027$ & $24.21$ \\[2pt]
\multirow{3}{*}{$10^3$}
 & DCP   & $49\,872$ & $33\,697$ & $67.57$ \\
 & DWD   & $865$     & $324$     & $37.46$ \\
 & \rev{DANS} & $11\,586$ & $4\,742$ & $40.93$ \\
\end{tabular}
\end{ruledtabular}
\caption{\label{tab:wimp_dynamic_rejection}
Rejected (unstable) DMACO configurations for $m_f=100~\mathrm{GeV}$. No stable DOY candidates were found.} 
\end{table}

\subsubsection*{Dark-oyster-like configurations}
\label{sec:doy_results}

For $m_f=10~\mathrm{GeV}$ and $y=10^3$, most sampled configurations satisfying the DOY conditions \rev{lie in the radially unstable part of the pressure map}. \rev{Their extended DM halo geometry does not guarantee stability}: the two-fluid pulsation calculation gives $\omega_0^2<0$ for these candidates.

For a geometric comparison with Ref.~\cite{Biesdorf2025}, we select stable $m_f=5~\mathrm{GeV}$ configurations in the independent ranges spanned by their Table V: \rev{$2.12\leq M/\Msun\leq7.98$, $0.60\leq R_{\OM}/\mathrm{km}\leq20.0$, and $22.8\leq R_{\DM}/\mathrm{km}\leq109.7$}. In the present sample, \rev{2839 of 2898 selected configurations (97.96\%)} satisfy $R_{\DM}/R_{\OM}>4$. This rectangular selection is broader than our working DOY definition and does not identify individual configurations. It is worth to stress that the comparison concerns small OM cores and extended DM halos, where the former have no explicit first-order hadron--quark transition as in Ref.~\cite{Biesdorf2025}.

\begin{figure*}[p]
\centering
\articlegraphic[width=\textwidth,height=0.90\textheight,keepaspectratio]{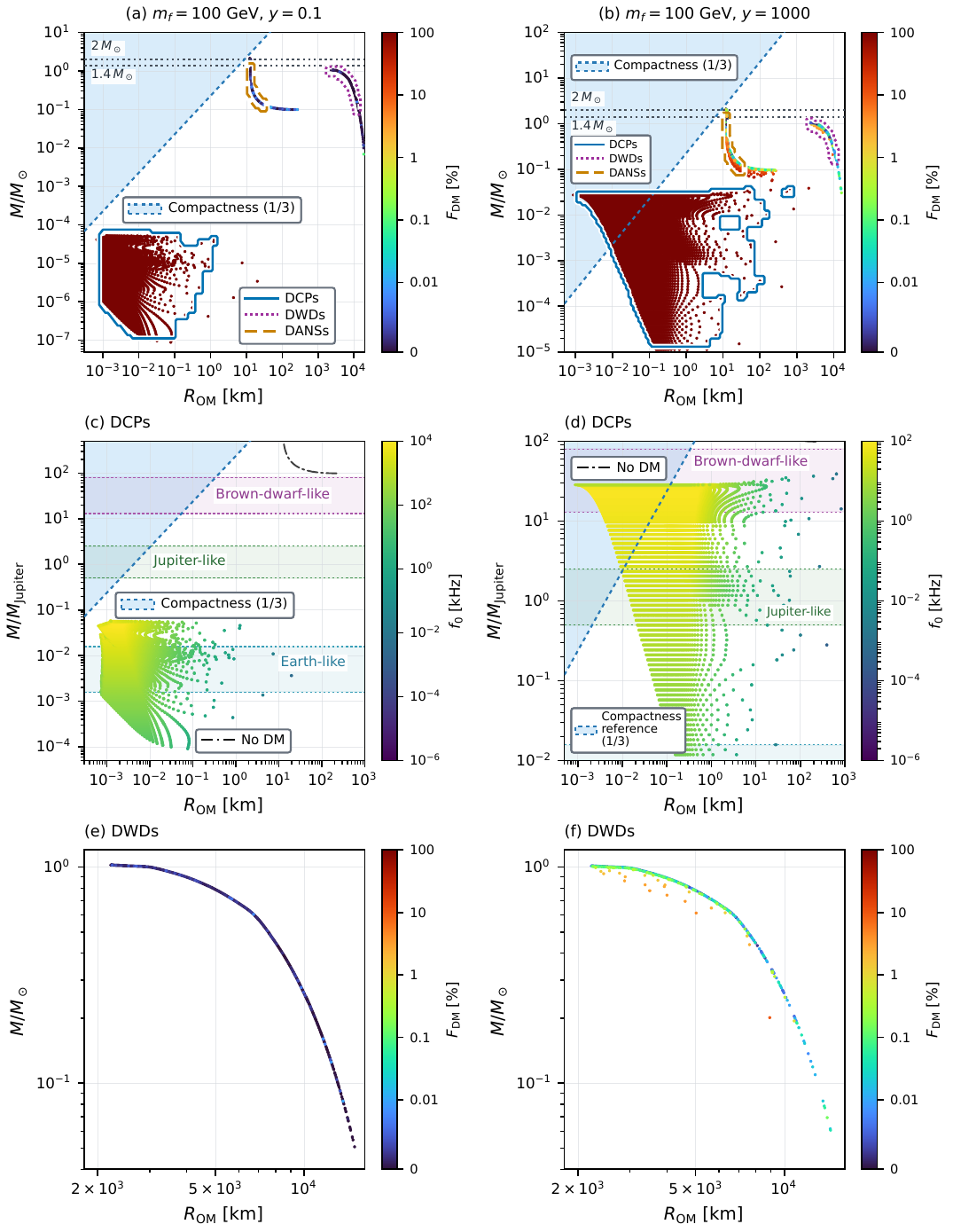}
\caption{\label{fig:wimpcombined}
Stable $m_f=100~\mathrm{GeV}$ configurations for $y=0.1$ (left panel) and $y=10^3$ (right panel). The first row shows global mass--radius distributions colored by $100F_{\DM}$; \rev{the second shows DCPs colored by $f_0$; the third shows DWDs colored by $100F_{\DM}$}. \rev{Axes match the corresponding lower-mass panels at fixed interaction strength.} DCP masses are in Jupiter masses, while the global and DWD panels use solar masses. Structural selections and reference curves have the meaning specified in Figs.~\ref{fig:mr_frequency} and \ref{fig:weak2}. The pressure-space stability calculations are retained in the analysis but are \rev{not repeated as separate figures}.}
\end{figure*}

\subsection{\texorpdfstring{Approaching WIMP masses: \\DMACOs with $m_f=100$~{\rm GeV}}{Approaching WIMP masses: DMACOs with mf = 100 GeV}}
\label{sec:wimp}

Weakly interacting massive particles (WIMPs) are DM candidates interacting gravitationally as well as weakly nongravitationally. They arise in several extensions of the standard model, with a broad, model-dependent mass range~\cite{Schumann:2019eaa}. Recent discussion of a single high-energy LUX-ZEPLIN event~\cite{Castelvecchi2026} provides additional motivation for exploring heavier dark particles. The collaboration reports a nuclear-recoil-like event that can be accommodated by several interaction models, with a global significance of $2.6\sigma$ and the largest local significances typically at WIMP masses above $200~\mathrm{GeV}$~\cite{LZ:2026axp}. This is not a confirmed DM detection or a model-independent lower mass bound. Our choice $m_f=100~\mathrm{GeV}$ is an independent benchmark for the compact-object calculation, not a fit to that event. Moreover, our two-fluid model retains gravity-only coupling between OM and DM; it does not model the nongravitational scattering probed by direct detection.

Therefore, we extend our calculation to $m_f=100~\mathrm{GeV}$ while retaining the same nonannihilating, self-interacting fermion model.
Figure~\rev{\ref{fig:wimpcombined}} compares the \rev{globally stable DCP and DWD} mass--radius diagrams for both \rev{self-interaction strengths}.
Table~\ref{tab:wimp_dynamic_rejection} reports the (unstable) rejected  DMACOs.

As occurred with previous results, the $100~\mathrm{GeV}$ results preserve the separation of oscillation-period scales found at lower masses, but push the DM-supported planetary configurations to smaller masses and radii.  The median DCP frequencies are about $2.01\times10^3~\mathrm{kHz}$ for $y=0.1$ and $72.9~\mathrm{kHz}$ for $y=10^3$. Thus, self-repulsion expands the dark configuration and lowers its frequency by more than an order of magnitude.  DWDs remain near $1.8\times10^{-4}~\mathrm{kHz}$ and \rev{DANSs} near $2.6~\mathrm{kHz}$ for both couplings because their OM structure still sets the dominant scale.  The large DCP rejection, $68$--$73\%$, shows that the equilibrium planetary window extends well beyond the dynamically stable branch at this fermion mass. No sampled $100~\mathrm{GeV}$ configuration satisfies the DOY conditions.

The weakly repulsive residual DMACOs have medians of $M=0.105\Msun$,
$R_{\OM}=54.1~\mathrm{km}$, and $F_{\DM}=2.33\times10^{-3}\%$, continuing
the intermediate-radius, OM-dominated family at $m_f=5$ and
$10~\mathrm{GeV}$. For strong repulsion, the residual family has medians of
$M=0.102\Msun$, $R_{\OM}=53.6~\mathrm{km}$,
$R_{\DM}=0.291~\mathrm{km}$, and $F_{\DM}=0.205\%$. Neither residual is
an extreme-halo or DOY configuration.

\subsection{Sensitivity of Stable DWDs to the Ordinary-Matter EoS}
\label{sec:dwd_eos_test}

In order to approach a more realistic description, we examine the sensitivity of structure and radial stability of DWDs to a different model for white-dwarf matter by comparing the adopted BPS-based low-pressure prescription with a cold electron--ion EoS \cite{Schaffner:2020}. In the corresponding one-fluid BPS calculation, the largest stable white-dwarf mass is about one solar mass. However, this maximal value should approach the 1.4 solar-mass value according to current models for white dwarfs \cite{Schaffner:2020}. It is therefore necessary to perform an additional study of DWDs but replacing the BPS EoS by the zero-temperature electron--ion EoS, where one adds to the electron energy density the ion rest-mass term $m_N\mu_e k_F^3/(3\pi^2)$, where $m_N$ is the nucleon mass and $k_F$ is the electron Fermi momentum, and the \rev{number of baryons per electron} is $\mu_e=A/Z=2$ for fully ionized \rev{$^4\mathrm{He}$}, $^{12}\mathrm{C}$, or $^{16}\mathrm{O}$. \rev{Composition-dependent electron captures and nuclear reactions are not included.} 

\clearpage
\onecolumngrid
\noindent\begin{minipage}{\textwidth}
\makeatletter\def\@captype{figure}\makeatother
\centering
\articlegraphic[width=\textwidth,height=0.52\textheight,keepaspectratio]{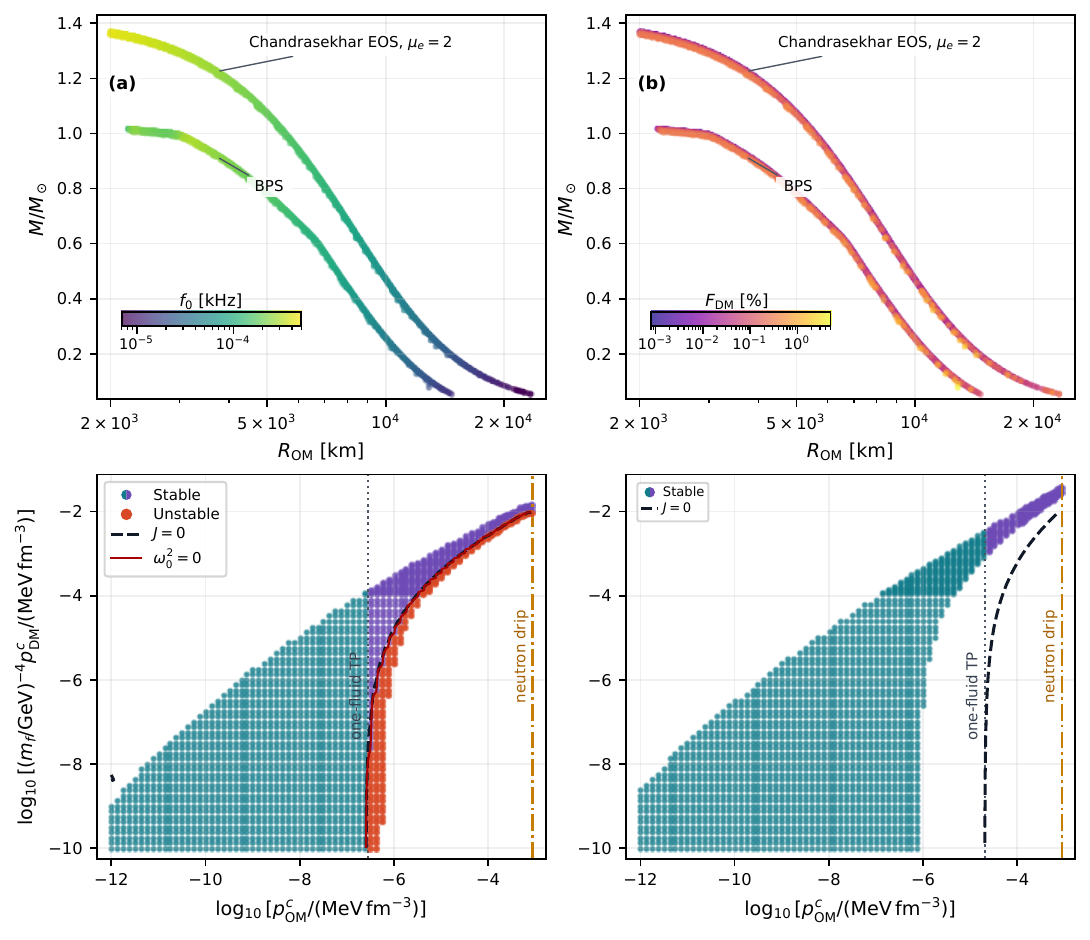}
\caption{\label{fig:dwd_eos_comparison}
TOV structure and radial stability of DWDs with $m_f=5~\mathrm{GeV}$ and $y=0.1$, using BPS and Chandrasekhar electron--ion matter with $\mu_e=2$~\cite{Schaffner:2020} for OM. Panels (a) and (b) show stable mass--radius configurations colored by $f_0$ and the DM mass percentage $100F_{\DM}$. The lower-left and lower-right panels show the BPS and Chandrasekhar pressure maps. Teal and purple points are stable; purple marks $p_{\OM}^c$ above the corresponding one-fluid turning point (TP). \rev{Red} points in the BPS panel are \rev{radially} unstable. In the lower-left panel, the dashed black $J=0$ and solid red $\omega_0^2=0$ curves coincide at the plotted resolution. In the lower-right panel, no red contour appears because $\omega_0^2>0$ throughout the sampled DWDs; \rev{the dashed $J=0$ locus} does not \rev{delimit a stable-to-unstable transition} within that \rev{selected DWD set}. Vertical dotted and dash-dotted lines mark the one-fluid TP and the OM pressure corresponding to neutron drip, respectively.}
\end{minipage}
\par\vspace{8pt}
\twocolumngrid

We refer to this zero-temperature electron--ion model as the \rev{Chandrasekhar EoS}~\cite{Schaffner:2020}.  In the past, the same EoS for WD matter was used in Ref.~\cite{Carvalho2025} to study the radial stability of DWDs, although in the one-fluid limit with a free DM fluid by fixing the central DM Fermi momentum. Besides, Ref.~\cite{Carvalho2026} explored the tidal deformability of two-fluid DWD configurations at fixed $q=M_{\DM}/M_{\OM}$, which is related to our $F_{\DM}=q/(1+q)$. Unlike their approach, we map the full coupled radial spectrum over a two-dimensional DWD central-pressure domain, rather than selecting fixed-$q$ sequences. We use $m_f=5~\mathrm{GeV}$ and $y=0.1$, for which the original scan produces the largest sampled DWD population. For both the BPS and Chandrasekhar EoS, we impose an upper-pressure cutoff in the OM fluid at the pressure corresponding to neutron drip for physical consistency \cite{Schaffner:2020}.

\begin{figure*}[t]
\centering
\articlegraphic[width=\textwidth]{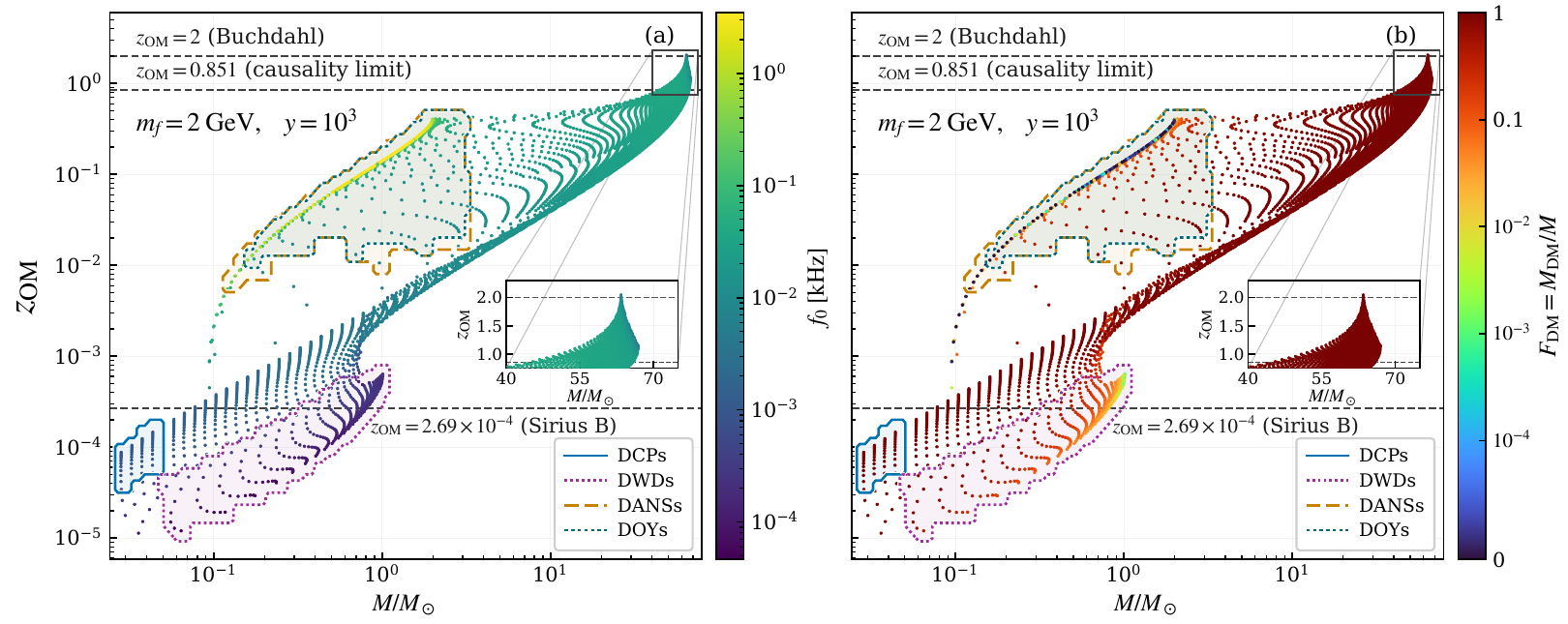}
\caption{\rev{OM-surface gravitational redshift, $z_{\rm OM}$, versus total mass for $m_f=2~\mathrm{GeV}$ and $y=10^3$. Colors indicate $f_0$ in kHz in in panel (a) and the dark mass fraction $F_{\DM}$ in panel (b). For comparison, we also display a dashed line for the Buchdahl  bound $z_{\rm OM}=2$. Additionally we put horizontal lines for the causal limit, $z_{\rm OM}=0.851$, and the one corresponding to Sirius~B white-dwarf $z_{OM}=2.69\times10^{-4}$. Notice that the outer-surface bounds do not directly apply to an OM emitting surface inside a DM halo. The insets magnify the radially stable configurations between the Buchdahl and causality bounds.}}
\label{fig:rs27_redshift}
\end{figure*}

In the one-fluid limit, we first found that the maximum mass increases from $1.012\Msun$ for BPS EoS to $1.394\Msun$ for the Chandrasekhar EoS, thus recovering the conventional scale. The stable DWD median mass increases from $0.790$ to $1.075\Msun$, and the largest selected stable mass from $1.012$ to $1.363\Msun$. These shifts are about $36\%$ and $35\%$, respectively, and therefore reflect a change due to the OM thermodynamics. Figure~\ref{fig:dwd_eos_comparison} shows the corresponding two-fluid
sequences and pressure maps when using both EoS while gravitationally interacting with DM.

Two-fluid radial oscillations remove $8.5\%$ of the sampled BPS DWD candidates, while no unstable configurations are found in the sampled Chandrasekhar sector. For each OM EoS, the scan includes central pressures \rev{on both sides of the turning point of the one-fluid white-dwarf sequence, within the adopted neutron-drip limit}. Stability of the admixed configurations is then determined from the coupled two-fluid equations. Among candidates above the corresponding one-fluid turning-point pressure, $72.5\%$ of the BPS sample and all sampled Chandrasekhar candidates remain radially stable. We highlight that the one-fluid turning point is \rev{shown only as a reference}.

\begingroup
\color{black}
\subsection{\rev{Gravitational Redshift at the OM Surface}}
\label{sec:surface_redshift}

A potentially relevant observable to measure our DMACOs are their gravitational redshifts, as occurs with one-fluid NSs \cite{Glendenning:2000}. However, in our present case of two-fluid stars, extended dark halo non-trivially change these gravitational redshift. One can again consider a photon emitted at $R_{\OM}$ and a static observer at infinity, while assuming that DM is transparent to light. For our Schwarzschild-like metric
{
\begin{equation}
 \dd s^2=-e^{2\Phi(r)}\dd t^2+
 \frac{\dd r^2}{1-2m(r)/r}+r^2\dd\Omega^2,
 \label{eq:rs27_metric}
\end{equation}}
with $\Phi(\infty)=0$, the frequency shift becomes \cite{Glendenning:2000}
{
\begin{equation}
 1+z_{\OM}\equiv\frac{\nu_{\rm em}}{\nu_\infty}
 =e^{-\Phi(R_{\OM})},
 \label{eq:rs27_definition}
\end{equation}}
where the normalization of $\Phi$ must be fixed at the outer radius $R=\max(R_{\OM},R_{\DM})$ since the spacetime becomes Schwarzschild, rather than at the OM surface when it lies inside a DM halo. The two-fluid TOV equations then give \cite{ShawqiMorsink2024}
{
\begin{align}
 1+z_{\OM}&=\frac{e^{\Delta\Phi_{\rm halo}}}{\sqrt{1-2M/R}},
 \label{eq:rs27_halo}\\
 \Delta\Phi_{\rm halo}&=\int_{R_{\OM}}^R
 \frac{m(r)+4\pi r^3p(r)}{r[r-2m(r)]}\,\dd r,
 \label{eq:rs27_integral}
\end{align}}
which depend on both fluids through $m$ and $p$. For a DM core $R=R_{\OM}$, so the integral vanishes and the usual surface formula is recovered \cite{Glendenning:2000}. Meanwhile, for a DM halo, it is positive: the fluid above the emitting surface contributes to its gravitational time dilation~\cite{ShawqiMorsink2024}.

Related halo-enhanced ordinary-surface redshifts have recently been investigated for hybrid stars with bosonic DM~\cite{Rather2026redshift} showing also $z_{\rm OM}>2$ being strongly related to the phase transition. Here we also evaluate this observable for our radially stable DMACO configurations which for some DM parameters display large DM halos.

Figure~\ref{fig:rs27_redshift} shows the full radially stable subset available for $m_f=2~\mathrm{GeV}$ and $y=10^3$, including DM-core and DM-halo configurations. We choose this parameter pair because it contains the largest dark-supported masses in our survey. Both panels show the same individual configurations separated by our compact classification, colored by $f_0$ and $F_{\DM}$, respectively. For a quantitative comparison with known bounds, we also display in this Fig. \ref{fig:rs27_redshift} the one-fluid causality bound  $z_{\rm OM}=0.851$ \cite{Schaffner:2020}, the redshift of the close white dwarfs Sirius B, $z_{\rm OM}=2.69\times 10^{-4}$ \cite{Joyce2018} and the Buchdahl limit, $z_{\rm OM}=2$ \cite{Schaffner:2020}. Notice that the last bound applies at the outer vacuum surface of an isotropic star with nonincreasing total energy density~\cite{Buchdahl1959redshift}. However, it should not be considered a bound on the redshift at an OM emitting surface inside a DM halo.

The largest sampled value is $z_{\OM}\simeq2.044$ (slightly surpassing the causality bound) for a radially stable configuration with $M\simeq63.30\Msun$, $R_{\OM}\simeq11.85~\mathrm{km}$ and $R_{\DM}\simeq451.05~\mathrm{km}$. This means that the residual configurations with high $z_{\rm OM}$ are mostly composed of DM (see panel (b)). Besides, DANSs and DOYs have typical NS values around 0.1. On the other hand, DWDs occupy the same range as typical white dwarfs like Sirius B, thus being difficult to be distinguished. Interestingly, DCPs have $z_{\rm OM}\sim 10^{-4}$ (and mostly composed by DM) similar to those of white dwarfs and, importantly, their masses are approximately two orders of magnitude lower, thus being attractive objects to be distinguished and searched in future astronomical surveys.
\endgroup

\section{Summary and final remarks}
\label{sec:conclusion}

We have calculated the fundamental-mode frequencies of two-fluid radial adiabatic oscillations in dark-matter-admixed compact objects (DMACOs). The OM fluid is modeled using EoS-I of Ref.~\cite{Annala2020}, while the dark component follows a self-interacting fermionic EoS with particle masses $m_f=2$, 5, 10, and $100~\mathrm{GeV}$ and strengths $y=0.1$ or $10^3$. \rev{In contrast to past work that relied on structural solutions, we determine radial stability from the lowest eigenvalue of the} coupled pulsation problem \rev{and compare it with static critical curves using the turning point criteria. This distinguishes boundaries of the stable domain from critical curves inside an already unstable region.}

A central result is the distinction between a static turning-point curve and the boundary of strict radial stability. In the low-pressure benchmark calculations, some $J=0$ curves correspond to \rev{higher radial modes reaching zero while the fundamental mode remains unstable}. This changes the number of unstable modes \rev{without restoring stability or contradicting the generalized turning-point criterion}~\cite{Chen2026}. Calculating the radial modes identifies which one reaches zero: the stable-domain boundary is associated with $\omega_0^2=0$, whereas \rev{a zero of a higher mode with $\omega_0^2<0$ leaves the configuration unstable}.

The frequency maps also reveal contrasting responses to DM microphysics. DWDs have typical fundamental-mode frequencies near $2\times10^{-4}~\mathrm{kHz}$, and \rev{DANSs} typically have frequencies of order $1$--$3~\mathrm{kHz}$. DCP frequencies depend much more strongly on the fermion mass and self-interaction strength. Increasing $m_f$ shifts dark-supported configurations toward smaller masses and radii and higher fundamental-mode frequencies. Strong self-repulsion expands the dark component and lowers the DCP frequencies, lengthening their oscillation periods. These trends reflect the contrasting structural roles of DM and OM\rev{. Each radial mode describes coupled motion of both fluids; their displacement profiles can differ although they share the same mode frequency.}

For $m_f=2$, 5, and $10~\mathrm{GeV}$, radial instability truncates the DWD and \rev{DANS} candidate windows at both \rev{self-interaction strengths}, whereas all sampled strongly repulsive DCP candidates remain stable. Strong repulsion also supports an extended DM-dominated family, with a largest sampled stable mass of $66.79\Msun$ for $m_f=2~\mathrm{GeV}$. The small OM contribution at its high-mass end and the $m_f^{-2}$ mass scaling connect this family to the self-interacting fermion-star limit~\cite{Narain2006}, rather than to ordinary NSs containing a small dark admixture. Their core--halo hierarchies include the extreme-halo and DOY subsets. At $m_f=100~\mathrm{GeV}$, stable members of all three primary classes remain, but most sampled DCP candidates are unstable and no sampled configuration satisfies the adopted DOY cuts.

We also replaced the OM BPS EoS with the Chandrasekhar electron--ion EoS to quantify whether it favors the existence of more stable DWD configurations. The latter EoS restores the conventional white-dwarf mass scale and broadens the sampled stable DWD domain. Importantly, stable two-fluid equilibria occur above the corresponding one-fluid turning point: $72.5\%$ of the sampled BPS candidates in that pressure range, and all corresponding sampled Chandrasekhar candidates, remain stable within the adopted sub-neutron-drip domain. \rev{The one fluid OM turning point is a reference for this comparison}, not \rev{the stability boundary of the two-fluid system}.

\rev{An extended DM halo also enhances the gravitational redshift at the OM surface. For $m_f=2~\mathrm{GeV}$ and $y=10^3$, the largest sampled value among radially stable configurations is $z_{\OM}\simeq2.044$. This redshift is evaluated at an emitting surface inside the halo, not at the outer vacuum boundary to which the usual Buchdahl surface applies.}

Thus, DM microphysics reshapes the stable mass, radius, and frequency domains. However, formation scenarios remain open questions for future research. From an observational perspective, a relevant DCP signature would be a planetary or substellar mass combined with an unusually small visible radius. Reference~\cite{Tolos2015} reported weakly interacting benchmark sequences with $M\sim10^{-6}$--$10^{-4}\Msun$ or below and ordinary radii of order meters, and strongly interacting sequences with $M\sim10^{-4}$--$10^{-2}\Msun$ or below and radii of order kilometers. Our two-dimensional survey extends beyond these sequences and populates the adopted Earth-, Jupiter-, and brown-dwarf-like mass windows over a broader range of $R_{\OM}$. 

Observations with the James Webb Space Telescope of IC~348 have identified objects with model-inferred masses near $2M_{\rm Jup}$~\cite{Luhman2025}. The sampled stable DCPs include the interval $1.8$--$2.2M_{\rm Jup}$ for $m_f=2,5,10~\mathrm{GeV}$ at $y=0.1$ and $m_f=5,10,100~\mathrm{GeV}$ at $y=10^3$, with DM mass percentages above \rev{$99.4\%$}. Unfortunately, this is only an overlap in mass scale, not an identification of the observed objects. Their mass estimates use evolutionary models without DM; a physical connection would require an appropriate formation and thermal-evolution calculation.

Finally, modern detection techniques measuring radial velocities, astrometry, and microlensing \cite{Vandorou2026} might serve to constrain masses, while transits supply radius information through the occulted stellar area~\cite{Perryman2000,LovisFischer2010}. These complementary measurements motivate future tests of unusually large mass-to-radius ratios, although assessing the reach of particular observatories requires a dedicated detectability analysis.

\begin{acknowledgments}
The authors thank Jürgen Schaffner-Bielich for fruitful conversations and helpful suggestions. J.C.J. is supported by Conselho Nacional de Desenvolvimento Científico e Tecnológico (CNPq) with Grants No. 151390/2024-0,  201218/2025-9, 201307/2026-0 and a grant from the APS-EPS-FECS-ICTP Travel Award Fellowship Programme (ATAP), Trieste, Italy. J.C.J. also acknowledges the kind hospitality of the Center for Nuclear Research at Kent State University, where part of this work was done. O.P. thanks Luciana for her moral support and constant encouragement. L.T. acknowledges support by the projects PID2022-139427NB-I00 and PID2025-168786NB-I00 financed by the Spanish MCIN/AEI/10.13039/501100011033/FEDER and ERDF/EU, and by the Grant CIPROM 2023/59 of Generalitat Valenciana.
\end{acknowledgments}

\end{document}